\documentclass[sigconf]{acmart}

\renewcommand\footnotetextcopyrightpermission[1]{} % removes footnote with conference information in first column

\usepackage{epsfig}
\usepackage{graphicx}
\usepackage{amsmath}
\usepackage{amssymb}

\usepackage{enumitem}
\usepackage{tablefootnote}
\usepackage[linesnumbered,titlenumbered,ruled,vlined,commentsnumbered,noend]{algorithm2e}
\usepackage{stfloats}
\usepackage{makecell}
\usepackage{colortbl}

\usepackage{balance}
\usepackage[rightcaption]{sidecap}

\definecolor{linen}{rgb}{0.96, 0.94, 0.93}
\definecolor{lightcyan}{rgb}{0.88, 1.0, 1.0}
\definecolor{lightyellow}{rgb}{1.0, 1.0, 0.88}
\definecolor{azure}{rgb}{0.94, 1.0, 1.0}

\usepackage[font=footnotesize,labelfont=bf]{caption}

\usepackage{multirow}
\usepackage{tabularx}
\usepackage{dsfont}
\usepackage{url}
\usepackage[tight]{subfigure}
\usepackage{color}
\usepackage{subfigure}
\usepackage{amsthm}
\usepackage[export]{adjustbox}
\usepackage{comment}

\usepackage[utf8]{inputenc} % allow utf-8 input
\usepackage[T1]{fontenc}    % use 8-bit T1 fonts
\usepackage{url}            % simple URL typesetting
\usepackage{booktabs}       % professional-quality tables
\usepackage{amsfonts}       % blackboard math symbols
\usepackage{nicefrac}       % compact symbols for 1/2, etc.
\usepackage{microtype}      % microtypography

\usepackage{balance}

\definecolor{dg}{rgb}{0,0.694,0.298}
\definecolor{purple}{rgb}{0.4,0.176,0.569}
\usepackage{pifont}% http://ctan.org/pkg/pifont
\usepackage{xspace}
\makeatletter
\DeclareRobustCommand\onedot{\futurelet\@let@token\@onedot}
\def\@onedot{\ifx\@let@token.\else.\null\fi\xspace}
\def\eg{\emph{e.g}\onedot} 
\def\ie{\emph{i.e}\onedot} 
 
\def\etc{\emph{etc}\onedot} 
 
\def\etal{\emph{et al}\onedot}
\makeatother

\AtBeginDocument{%
  \providecommand\BibTeX{{%
    \normalfont B\kern-0.5em{\scshape i\kern-0.25em b}\kern-0.8em\TeX}}}

\begin{document}
\fancyhead{}

% \title{Hearing isn't Believing: I Know Who is Real Alice}
% \title{\emph{DeepSonar}: Towards Effective and Robust AI-Synthesized\\Fake Voices Detection}
\title{\emph{DeepSonar}: Towards Effective and Robust Detection of\\AI-Synthesized Fake Voices} % Felix

\author{Run Wang$^{1}$, \ Felix Juefei-Xu$^{2}$, \ Yihao Huang$^{3}$, \ Qing Guo$^{1,\dagger}$, \ Xiaofei Xie$^{1}$, \ Lei Ma$^{4}$,  \ Yang Liu$^{1,5}$}
\thanks{
% Run Wang's email: wangrun@whu.edu.cn \\
Run Wang's email: runwang1991@gmail.com \\
$^{\dagger}$ Qing Guo is the corresponding author~({tsingqguo@gmail.com})}
\affiliation{\institution{$^{1}$Nanyang Technological University, Singapore  \ \ $^{2}$Alibaba Group, USA}}
\affiliation{\institution{$^{3}$East China Normal University, China \ \ $^{4}$Kyushu University, Japan}}
\affiliation{\institution{$^{5}$Institute of Computing Innovation, Zhejiang University, China}}
% \affiliation{\institution{wangrun@whu.edu.cn, \ \ \{juefei.xu, huangyihao22\}@gmail.com, \ \ \{qing.guo, xfxie\}@ntu.edu.sg, \\ malei@ait.kyushu-u.ac.jp, yangliu@ntu.edu.sg}}

\renewcommand{\shortauthors}{Run Wang, et al.}
\renewcommand{\authors}{Run Wang, Felix Juefei-Xu, Yihao Huang, Qing Guo, Xiaofei Xie, Lei Ma, Yang Liu}

\begin{abstract}
% DeepFakes are real potential threats to everyone including politicians, celebrities, and common people. Existing studies are mostly focused on detecting manipulated facial images, which is largely not enough for fighting against DeepFakes since synthesized fake voices are common in DeepFakes and not fully tackled. Voices could be synthesized more indistinguishable than ever before with the rapidly development of various AI techniques such as WaveNet\footnote{https://deepmind.com/blog/article/wavenet-generative-model-raw-audio}. Image and voice synthesis will work together to produce high-quality DeepFake videos, which will bring challenges to detectors and pose severe threats to community.
%  which are real potential threats to everyone including politicians, celebrities, and common people

With the recent advances in voice synthesis, AI-synthesized fake voices are indistinguishable to human ears and widely are applied to produce realistic and natural DeepFakes, exhibiting real threats to our society. However, effective and robust detectors for synthesized fake voices are still in their infancy and are not ready to fully tackle this emerging threat. In this paper, we devise a novel approach, named \emph{DeepSonar}, based on monitoring neuron behaviors of speaker recognition (SR) system, \ie, a deep neural network (DNN), to discern AI-synthesized fake voices. Layer-wise neuron behaviors provide an important insight to meticulously catch the differences among inputs, which are widely employed for building safety, robust, and interpretable DNNs. In this work, we leverage the power of layer-wise neuron activation patterns with a conjecture that they can capture the subtle differences between real and AI-synthesized fake voices, in providing a cleaner signal to classifiers than raw inputs. Experiments are conducted on three datasets (including commercial products from Google, Baidu, \etc) containing both English and Chinese languages to corroborate the high detection rates (98.1\% average accuracy) and low false alarm rates (about 2\% error rate) of DeepSonar in discerning fake voices. Furthermore, extensive experimental results also demonstrate its robustness against manipulation attacks (\eg, voice conversion and additive real-world noises). Our work further poses a new insight into adopting neuron behaviors for effective and robust AI aided multimedia fakes forensics as an inside-out approach instead of being motivated and swayed by various artifacts introduced in synthesizing fakes.

\begin{figure}[t]
	\centering
	\includegraphics[width=\columnwidth]{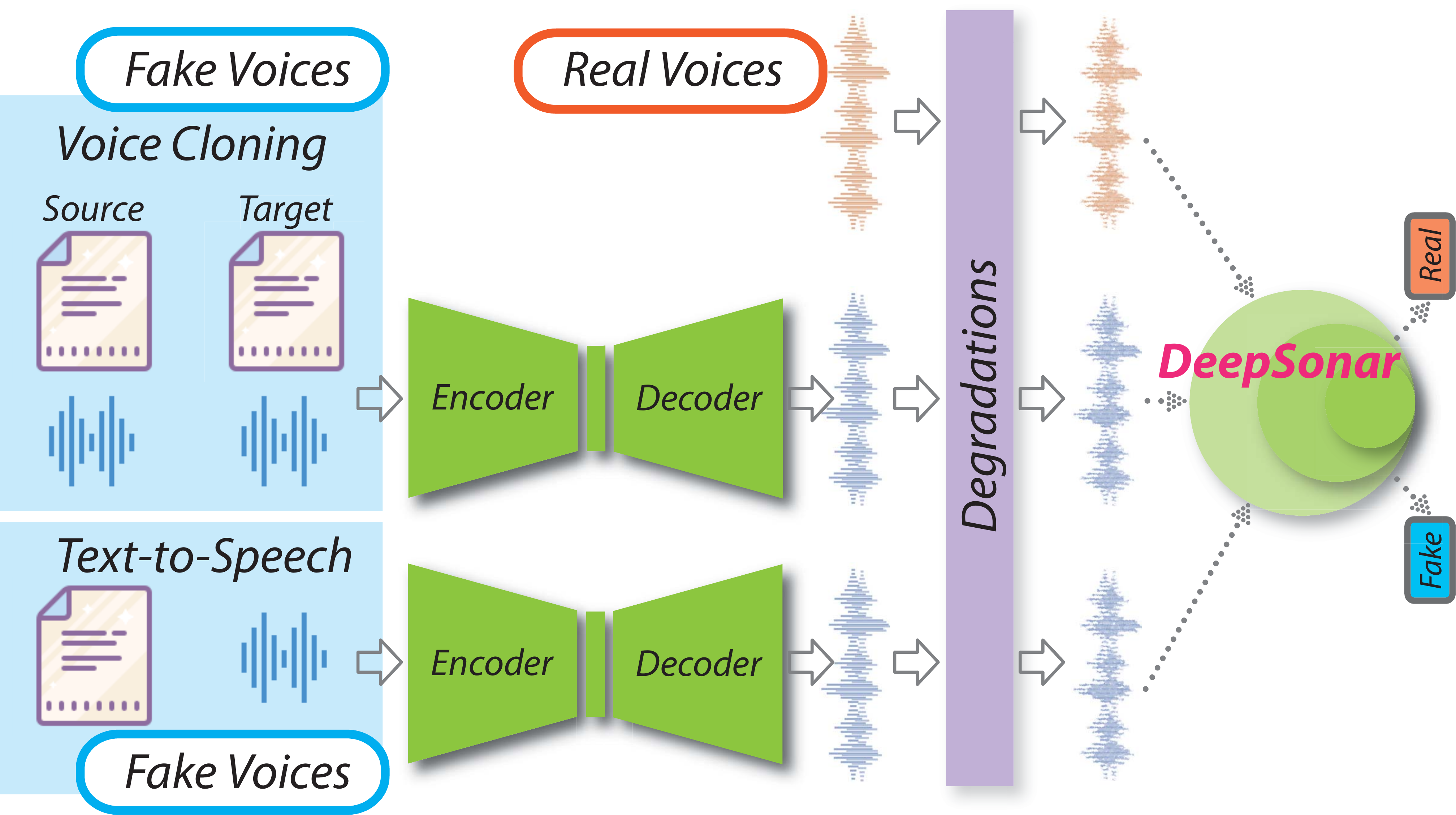}
	\caption{Two types of fake voices, voice cloning and text-to-speech. Voice cloning is more likely a voice style transfer by giving a "source voice" and output a cloned style similar "synthesized voice". Text-to-speech can generate a new voices by any given texts having specific timbre. Degradation indicates our proposed approach DeepSonar can handle voices that are manipulated by voice conversions and additive real-world noises.}
% 	RA captures real human voices in physical environments and re-presented with a replay device, but the content is the original one rather than regenerated, like TTS and VC.
	\label{Figure:fig1}
	\vspace{-10pt}
\end{figure}
%------------------------
\end{abstract}




% %   \includegraphics[width=\textwidth]{sampleteaser}

\maketitle

% symbols
% \etal \eg \ie \etc

%-----------------------------------------------------------------------
%-----------------------------------------------------------------------
%------------------------
\section{Introduction}\label{sec:intro}

In August 2019, the wall street journal reported the news titled "\emph{Fraudsters Used AI to Mimic CEO’s Voice in Unusual Cybercrime Case}" \cite{report}. In this report, criminals used AI-based software to impersonate a CEO's voice and successfully swindled more than \$243,000 by speaking on the phone. Recently, advances in AI-synthesized techniques have shown its powerful capabilities in creating highly realistically sounded voices \cite{gu2018multi,shen2018natural}, indistinguishable images \cite{karras2019analyzing,wang2019fakespotter,huang2020fakelocator}, and natural videos \cite{chan2019everybody,suwajanakorn2017synthesizing,jiang2020deeperforensics10}. Human eyes and ears could be easily fooled by these realistic DeepFakes \cite{fake_image,fake_audio}. Furthermore, producing DeepFakes is easy with tools like FaceApp, ZAO, \etc. Thus, it also raises security and privacy concerns to everyone while we are enjoying the fun of these synthesized fakes. Powerful detection and defense mechanisms should be developed by the community for fighting against such DeepFakes \cite{tolosana2020deepfakes,mirsky2020creation}.
% In the early April 2018, Barack Obama called president Trump a ``Complete Dipsh*t'' in an online video\footnote{\url{https://youtu.be/cQ54GDm1eL0?t=23}}. Obviously, this is an AI-generated fake video (also known as DeepFakes \cite{tolosana2020deepfakes,mirsky2020creation}) powered by image and voice synthesis techniques.

Voice/speech synthesis steps into a new era since DeepMind developed WaveNet \cite{oord2016wavenet,wavenet} that could generate realistic and convincing voices. Improving the interaction experiences between machines and humans is the initial idea for developing voice synthesis techniques. Based on this idea, some commercial products like intelligent customer service are created by using voice synthesis techniques. Unfortunately, some attackers and criminals misuse them for illegal purposes like a politician giving an unreal statement, which may cause a regional crisis or someone imitating the victim's voice for fraud intentions. All of these can be easily performed without any effort by merely giving texts and a clip of the victim's real voice using some open-sourced tools \cite{kobayashi2018sprocket} or commercially available text-to-speech (TTS) systems. Thus, discerning whether a clip of voice is synthesized with AI or spoken by humans is extremely important in this era when hearing is not believing anymore.

TTS synthesis, voice cloning (VC), and replay attack (RA) are the three different modalities for synthesizing fake voices \cite{NAP25488}. TTS and VC involve the content regeneration, thus they are more realistic than RA and are difficult for human ears to distinguish. Therefore, they are especially worrisome and pose high risks. Figure \ref{Figure:fig1} shows a more detailed description of the two
types of fake voices. Recently, AI-synthesized fake voices have already drawn attention from the community. Google launched a challenge competition dedicated to spoofed voice detection \cite{Todisco2019}. Farid \etal proposed the first bispectral analysis method to distinguish human voices and AI-synthesized voices based on the observation of the bispectral artifacts in fake voices \cite{albadawy2019detecting}. However, existing works on discerning AI-synthesized fake voices all failed in fully tackling the aforementioned TTS and VC fake voices and thoroughly evaluating their robustness against manipulation attacks, which is extremely important for a detector deployed in the wild. Here, manipulation attacks indicate that the voices are corrupted with real-world noises (\eg, rain, laughing) or converted by manipulating their signals without altering its linguistic contents, such as resampling, and shifting of pitch.
% and released a dataset where voices are synthesized with outdated techniques rather than the state-of-the-art (SOTA) techniques like some commercial products adopted
% Farid \etal proposed the first bispectral analysis method for distinguishing real human voices and synthesized voices \cite{albadawy2019detecting}.

Voice synthesis and image synthesis are regularly combined for producing audio-visual consistent video DeepFakes. Compared to image synthesis, voice synthesis exhibits some differences and brings new challenges to detection. Firstly, artifacts in fake voices could be hardly sounded and provide sufficient clues for forensics. They are vastly different from artifacts in fake images that are easily noticed by eyes. Secondly, voice signals are one dimension signals. It is not as simple to introduce artifacts into the voice synthesis procedure as in images that have multiple channels spanning two dimensions spatially. Lastly, for voices recorded indoors or outdoors where noises are abundant, it is easy for the attackers to fool the detectors by adding real-world noises in such circumstances, thus robustness is essential for fake voice detectors.

% In recent years, a lot of research are working on detecting images synthesized by AI techniques mostly GANs \cite{wang2019fakespotter,wang2019cnn,stehouwer2019detection}, but studies on fake voices detection is rare. However, DeepFake often combines either image synthesis or voice synthesis (or both), which is claimed in a DeepFake detection challenge (DFDC) announced by Facebook. In comparing with images, voice signals are one dimension signals which exhibit significant differences to images \cite{lyu2020deepfake}. Moreover, artifacts in fake voice could be hardly sounded and provide sufficient clues for forensics, which is vastly different to artifacts in fake images that is easily noticed by eyes. Unfortunately, previous research on detecting audio spoofing could be hardly employed for explicitly addressing synthesized fake voices, because the content generation is ignored in these approaches \cite{zakariah2018digital}. AI-synthesized fake voices detection has already drawn attentions from both academia and industry. Farid \etal proposed a bispectral analysis method for distinguishing a real human voice from synthesized voice \cite{albadawy2019detecting}. Google launched a challenge competition dedicated to fake voices detection \cite{Todisco2019}.

In this paper, we propose a novel approach, named DeepSonar\footnote{Sonar is known as its powerful capabilities in sniffing and probing electronic devices underwater based on sound signals. We hope that our approach is a sonar in discerning AI-synthesized fake voices.} as presented in Figure \ref{Figure:fig1}, based on monitoring neuron behaviors of a DNN-based SR system with a simple binary-classifier to discern AI-synthesized fake voices. We conjecture that the layer-by-layer neuron behaviors in DNNs could provide more subtle features and cleaner signals for the classifiers than raw voice inputs, which served as an important asset for differentiating real human voices and fake voices. In this work, we are dedicated to the \textbf{TTS} and \textbf{VC} fake voices since they are AI-synthesized with content regenerated that are more indistinguishable than RA to our ears. To the best of our knowledge, this is the first work employing layer-wise neuron behaviors to discern AI-synthesized voices and conducting a comprehensive evaluation on its robustness against two manipulation attacks, 1) voice conversions, and 2) additive real-world noises.

% The third-party DNN is used for learning speech representations to recognize speakers.
% and reveal the differences between real and fake voices

% \begin{enumerate}[leftmargin=*]
%     \item \textbf{Effectiveness}. The detector should be general to different languages, genders, accents, and synthetic techniques.
%     \item \textbf{Robustness}. Voices could be recorded indoor or outdoor where real-world noises (\eg{}, rain, laughing) are abundant. Additionally, voice conversions like resampling are also common in post-processings for publishing purposes. A robust detector should be well tackling these manipulations.
%     % in voices as the speaking environments are always not clean and attacker will add noises for fooling detectors, which requires detectors dealing with noises well.
% \end{enumerate}

To comprehensively evaluate the effectiveness and robustness of our approach in discerning AI-synthesized fake voices, our experiments are conducted on three datasets including publicly available datasets, in which voices are synthesized with commercial products and self-built dataset with available open-sourced tools. In the experiments, we aim to evaluate the \textbf{effectiveness} of DeepSonar in distinguishing fake voices synthesized with different languages, synthetic techniques, \etc, and investigate the \textbf{robustness} of DeepSonar in tackling two manipulation attacks (including \textbf{voice conversion} and \textbf{additive real-world noises}). Experimental results have demonstrated that DeepSonar gives an average accuracy higher than $98.1\%$ and an equal error rate (EER) lower than $2\%$ on the three datasets. DeepSonar also outperforms prior work leveraging bispectral artifacts to differentiate fake voices \cite{albadawy2019detecting} in both effectiveness and robustness. Our main contributions are summarized as follows.% DeepSonar also outperforms prior work by using bispectral analysis technique to distinguish fake voices \cite{albadawy2019detecting} and a simple baseline with temporal convolution approach \cite{temporal_audio,Atlas_OSS}.
% and investigating mel-frequency spectrograms as input features with a DNN model to classify human speeches and fake voices \cite{temporal_audio,Atlas_OSS}

\begin{itemize}[leftmargin=*]
    \item \textbf{New observation of layer-wise neuron behaviors for discerning fake voices.} We observe that the layer-wise neuron behaviors capture more subtle features that provide cleaner signals for the classifiers than raw voice inputs for building effective and robust fake detectors. Thus, we propose DeepSonar based on this observation by monitoring neuron behaviors to reveal the differences between real voices and AI-synthesized fake voices.
    % \item \textbf{New observation of layer-wise neuron behaviors for discerning fake voices and new insight for fighting against DeepFakes.} We observe that the layer-by-layer neuron behaviors capture more subtle features than raw inputs which provides cleaner signals for classifier to build effective and robust fake detectors. Thus, we propose DeepSonar based on this observation by monitoring layer-wise neuron behaviors to reveal the differences between real human speeches and AI-synthesized fake voices. Furthermore, instead of handcrafted features or raw inputs with well-designed models for detecting specific fakes, our approach presents a new insight by leveraging the power of layer-wise neuron behaviors for differentiating real and fake in a general manner without sacrificing performances.
    \item \textbf{Performing a comprehensive evaluation of the effectiveness and robustness against manipulations attacks.} Experiments are conducted on three datasets where voices are synthesized with various techniques, containing English and mandarin Chinese languages spoken by males and females with different accents. Experimental results illustrated its effectiveness in discerning fake voices and robustness against two manipulation attacks, voice conversions and additive real-world noises.

    \item \textbf{New insights for fighting against AI aided multimedia fakes.} Instead of investigating the artifacts introduced by various synthetic techniques, our approach presents a new insight by leveraging the power of layer-wise neuron behaviors for differentiating real and fake in a generic manner. Furthermore, it also demonstrates the potentials for building robust detectors and evasion attacks, which are important to be deployed in the wild.
    \end{itemize}

% The remainder of our paper is organized as follows. Section \ref{sec:related} discusses the related work. Section \ref{sec:method} presents our method on discerning fake voices in detail. In Section \ref{sec:settings}, we introduce the experimental settings (\eg{}, datasets, baselines, and evaluation metrics) and implementation details. We show the experimental results and demonstrate the effectiveness and robustness of our proposed method in Section \ref{sec:exp_results}. Finally, Section \ref{sec:conc} concludes this paper and discusses the future research directions.

%-----------------------------------------------------------------------
%-----------------------------------------------------------------------
\section{Related Work}\label{sec:related}
% In recent years, we have witnessed many research efforts and encouraging progress on voice synthesis. This section briefly reviews existing work on speech synthesis and countermeasures in detecting synthesized fake voices.

\subsection{Voice Synthesis}

Voice synthesis can be divided into two categories: 1) non-DNN based, such as using hidden Markov models (HMMs) and Gaussian mixture models (GMMs) to learn speech features and replicate them, and 2) DNN based for synthesizing naturalness speech and even on unseen words.

The first technique is speech concatenation that concatenates some pre-recorded speech segments to synthesize a new clip voice \cite{yuan2019adversarial}. The other technique on format analysis uses acoustic models without a human voice as input to generate robotic-sounding speech \cite{taylor2009text}. Modeling the human vocal tract and vocal biomechanics is another technique for synthesizing speech, which is known as articulatory speech synthesis \cite{lucero2013physics}. Some studies explore leveraging HMM to modulate speech proprietaries like fundamental frequency and duration \cite{zen2007hmm}. These techniques are widely employed in the early years for speech synthesis, but suffer from naturalness issues, which could be easily sounded by human ears.

\textbf{DNN based}. DNN-based speech synthesis techniques directly map linguistic features to acoustic features by leveraging the power of DNNs in representation. Various models (\eg, Boltzmann machines \cite{ling2013modeling}, deep belief network \cite{kang2013multi}, mixed density networks \cite{bishop1994mixture}, Bidirectional LSTM \cite{li2017multi}) are proposed based on DNNs for synthesizing high quality and natural speech. Some synthesized samples are available online \cite{audio_sample}.

% Sequence-to-sequence (seq2seq) neural networks could be used for dealing the different length between input sequence and output sequence and employed in modeling acoustic features for speech synthesis \cite{wang2016first,zhang2018forward}.

WaveNet \cite{oord2016wavenet} developed by DeepMind in 2016 and Tacotron \cite{wang2017tacotron} created by Google in 2017 are two milestones in speech synthesis. The two models significantly promote the progress of speech synthesis, which enables large scale commercial applications for building TTS and VC systems. WaveNet originates from PixelCNN \cite{van2016conditional} or PixelRNN \cite{oord2016pixel} and shown its powerful capabilities in modeling waveforms with a generative model that is trained on a real audio dataset. Tacotron \cite{wang2017tacotron} is an end-to-end speech synthesis model that can be trained on <text, audio> pairs to avoid large human annotation efforts.
% Deep Voice2 \cite{gibiansky2017deep} combined Tacotron and WaveNet and achieved amazing performance in speech synthesis.
% Tacotron also uses generative model like WaveNet, but it utilizes sequence-to-sequence (seq2seq) model for mapping text to a spectrogram. Seq-2-seq network is designed for addressing the imbalance between input and output sequence in speech synthesis \cite{wang2016first,zhang2018forward}.
% However, WaveNet suffers low efficiency due to sample-level autoregressive nature and depends on the linguistic features of an existing TTS front-end. Parallel wavenet\cite{oord2017parallel} and Deep Voice \cite{arik2017deep} are proposed to address these issues.
%
Due to the powerful capabilities of WaveNet and Tacotron, some commercial products are developed based on them, such as Baidu TTS \cite{baidu}, Amazon AWS Polly \cite{amazon}, and Google Cloud TTS \cite{google}. Unfortunately, some attackers can maliciously use speech synthesis techniques and develop fake voices for fraud intentions, bringing potential security concerns.

\vspace{-20pt}
\subsection{Fake Voice Detection}

% Digital audio forensics and synthetic speech detection with DNNs are two major potential methods for detecting fake voices.
% \textbf{Digital audio forensics.}

In the past decades, some digital audio forensic studies are working on detecting various forms of audio spoofing \cite{zakariah2018digital}. These approaches examine metadata of audio files and investigate their actual bytes. Douglas \etal{} \cite{koenig2012forensic} examine the eleven audio recordings from three Olympus recorders in the digital header data for audio authentication. Malik \etal{} \cite{zhao2013audio} propose using acoustic environment signature as an important feature for detecting audio forgery by verifying the integrity of digital audio. These studies failed in addressing audio content that is synthesized.
% Container-based and content-based are two authentication method for digital audio forensics. Container-based analysis techniques examine file structure and metadata of audio file \cite{koenig2012forensic}, while content-based analysis techniques investigate actual bits and bytes of the audio files \cite{zhao2013audio}.

The most similar work to ours is \cite{albadawy2019detecting} that is the first study dedicated to AI-synthesized fake voices. In their work, they propose a bispectral analysis method for detecting AI-synthesized fake voices. They observe that specific and unusual spectral correlation exhibited in the fake voices synthesized with DNNs, which are called bispectral artifacts. Thus, they explore to use higher-order polyspectral features for discriminating fake voices. This work is also motivated by investigating artifacts introduced in fake voices like some recent studies on detecting fake images \cite{yang2019exposing,li2018exposing}. Artifact-based detectors will be invalid when the artifacts are fixed with some optimization methods or new synthetic techniques are proposed.

% The most similar work to us is \cite{albadawy2019detecting}. They proposed a bispectral analysis method for detecting AI-synthesized fake voices.
% \textbf{Synthetic speech detection with DNNs.} Yu \etal{} classified real and fake voices by extracting dynamic acoustic features \cite{yu2017spoofing}. Zhang \etal proposed using CNNs in conjunction to RNNs for differentiating synthetic speech \cite{zhang2017investigation}. Tian \etal{} proposed using temporal convolutional networks to discern fake voices rather than handcrafted features with traditional machine learning models \cite{tian2016spoofing}. These techniques are mostly applied for speaker recognition and verification with prior knowledge of specific speakers' real voices. Furthermore, these approaches are susceptible to additive real-world noises and suffer manipulation attack issues, which is not practical to be deployed in the wild.
% and examining fine-grained handcrafted features along with well-designed complex DNN models

In this paper, instead of investigating the artifacts in raw voices introduced in synthesis, we explore a new way by monitoring neuron behaviors of DNN-based SR systems with a simple binary-classifier to distinguish real and fake voices. The layer-wise neuron behaviors can capture more subtle features in differentiating real and fake voices. Experimental results show that our approach outperforms previous work (by investigating bispectral artifacts \cite{albadawy2019detecting}) in terms of both effectiveness and robustness.
%------------------------
\begin{figure}
\centering
\includegraphics[width=\columnwidth]{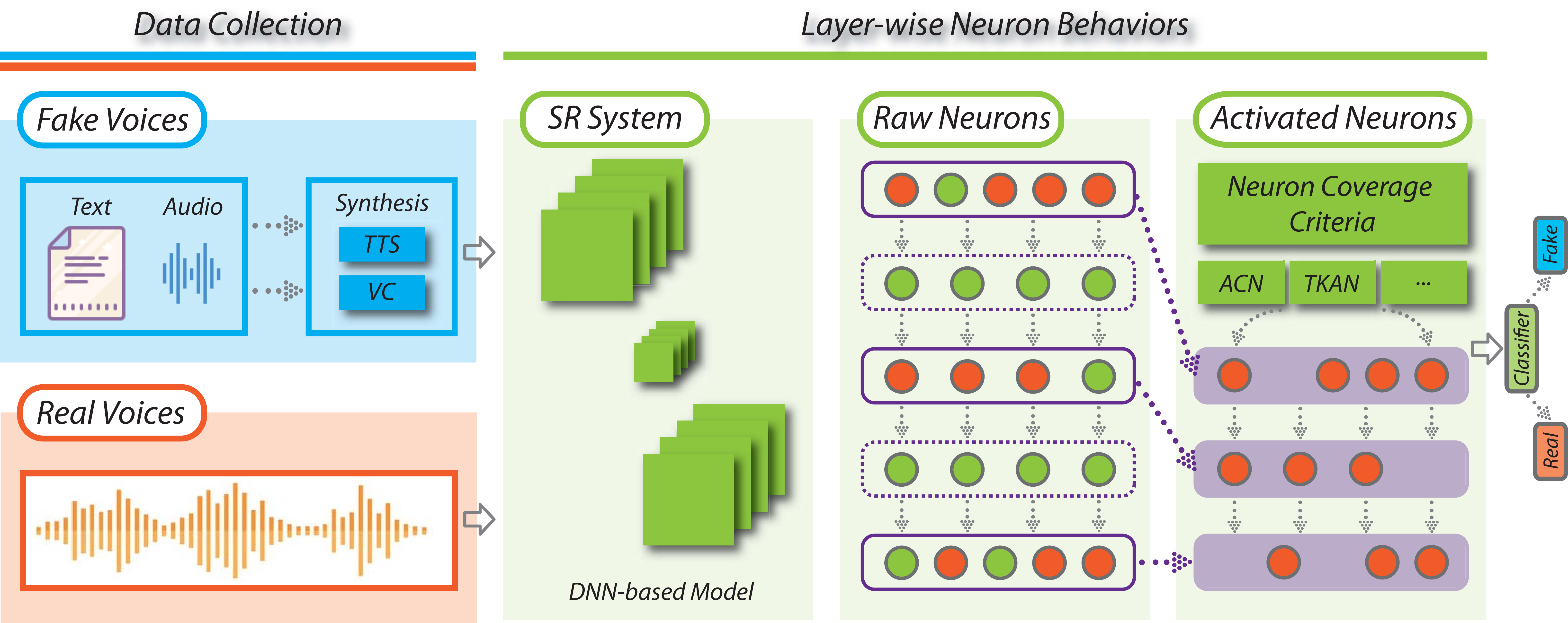}
\caption{The framework of our proposed DeepSonar. We collect numerous real human speeches and fake voices synthesized with VC and TTS techniques as inputs, then a DNN-based SR system is adopted to capture the raw layer-wise neuron behaviors of inputs and the designed neuron coverage criteria (\eg, ACN and TKAN) is employed to determine the activated neurons which are more valuable in hunting the subtle differences between inputs, finally a binary-classifier is trained based on the activated layer-wise behaviors of inputs to predict if a clip of voice is real or fake.}
\label{Figure:overview}
\vspace{-10pt}
\end{figure}
%------------------------

%-----------------------------------------------------------------------
%-----------------------------------------------------------------------
\section{Method}\label{sec:method}

We first introduce our basic insight in discerning fake voices, and then present the overview framework of DeepSonar, after which we detail how to capture the layer-wise neuron behaviors and detect fake voices with binary-classifier in the following subsections.

% introduce the two essential components of our designed architecture in the following subsections.

\subsection{Insight}

Monitoring neuron behaviors is an important technique for hunting the differences among a set of inputs to DNNs and investigating the internal behaviors of DNNs, which is widely employed in assuring the quality of DNNs \cite{pei2017deepxplore,ma2018deepgauge,odena2019tensorfuzz,xie2019deephunter}, protecting the safety of DNNs like fighting adversarial examples attack \cite{ma2019nic,ma2018mode}, and providing interpretation for DNNs \cite{tao2018attacks}, \etc{}

For quality assurance of DNNs, both DeepXplore \cite{pei2017deepxplore} and DeepGauge \cite{ma2018deepgauge} introduce neuron coverage as testing criteria to explore the amount of DNN logic covered by given a set of inputs. Neuron coverage is similar to code coverage in traditional software testing and used to explore the vulnerabilities of DNNs, which are susceptible to adversarial examples \cite{goodfellow2014explaining}. In ensuring the safety of DNNs, NIC \cite{ma2019nic} and MODE \cite{ma2018mode} exploit the critical neurons in DNNs for detecting adversarial examples and fixing issues that lead to misclassification in DNNs. In providing interpretation for DNNs, AMI \cite{tao2018attacks} explores the correlation between important neurons and human perceptible face attributes. Furthermore, the visualization techniques are also proposed \cite{olah2017feature, carter2019activation} to facilitate the understanding on the roles of neurons.

% \wang{add a figure to illustrate the capabilities of layer-wise neuron behaviors in hunting the minor differences between inputs.}

According to recent studies, neuron behaviors have demonstrated their powerful capabilities in investigating the internal behaviors of DNNs and revealing the minor differences among inputs like adversarial examples and legitimate inputs. In this work, we conjecture that layer-wise neuron behaviors could capture more subtle features and produce cleaner signals to a classifier than raw voice inputs in distinguishing the differences between inputs. Thus, we propose DeepSonar by monitoring layer-wise neuron behaviors of the DNN-based SR system with a simple binary-classifier to discern human speeches and AI-synthesized fake voices.

% Due to the mysterious power of neurons in DNN models, we explore whether neuron behaviors could be served as an asset in discerning fake voices as the intrinsic characteristic between real and fake voices is different, which could be captured by monitoring layer-by-layer neuron behaviors.

\subsection{Overview of \emph{DeepSonar} Framework}

We present the overview of DeepSonar framework in Figure \ref{Figure:overview}. In general, we first collect numerous real and synthesized fake voices with good diversity in languages, accents, genders, and synthetic techniques. Real voices are collected from public datasets and available free videos from the internet, which are spoken by humans in different languages, accents by males or females. In fake voice collection, we 1) use TTS techniques to synthesize new voices with merely given texts, and 2) utilize VC techniques to produce a clip of fake voices having similar timbre to real voices. Then, we adopt a DNN-based SR system to capture the layer-wise neuron behaviors for both real and fake voices and determine the activated neurons with designed neuron coverage criteria. Finally, the captured neuron behaviors are formed as input feature vectors for training a simple supervised binary-classifier based on shallow neural networks to predict whether a clip of voice is a human speech or synthesized.

\subsection{Layer-wise Neuron Behaviors}

% We propose a novel fake voice detector based on monitoring layer-wise neuron behaviors and the framework is shown in Figure \ref{Figure:overview}. In this work, we explore two different layer-wise neuron behaviors based on the strategies for determining activated neurons to evaluate their effectiveness in discerning real and fake voices.
% We employ a DNN-based SR system to capture the layer-wise neuron behaviors of voices as shown in Figure \ref{Figure:overview}.

Layer and neuron are the basic components in a DNN model. Each layer in a DNN has its own distinct role in learning the input representations \cite{mahendran2015understanding}. A neuron $x$ is the basic unit for representing the inputs in each layer, whose output is calculated by the activation function $\varphi$, previous layer neurons $X^{'}$, weights matrix $W$, and bias $b$, \ie, $\varphi(W\cdot X^{'}+b)$.

Neurons can be classified as activated neurons and inactivated neurons on a given input, according to recent studies in DNN testing \cite{pei2017deepxplore}. Here, an activated neuron means that its output value is large than a predefined threshold $\delta$, and vice versa. According to recent studies, activated neurons could carry more information than inactivate neurons and have a large influence on its following consecutive layers \cite{pei2017deepxplore,ma2018deepgauge,ma2019deepct,ma2018deepmutation}. Thus, we monitor the activated neurons to discern the differences among inputs.

% The first one is counting the number of activated neurons in each layer. The other one is selecting neurons having top K output values in each layer. Multiple layers facilitate the powerful capabilities of DNNs in learning various intermediate representations of inputs. However,

In monitoring the layer-wise neuron behaviors, we need to address the following three issues, 1) which DNN-based model is more suitable for monitoring neuron behaviors? 2) which layers in the model are elected to monitor neuron behaviors? 3) how to determine the threshold $\delta$ using neuron coverage criteria?

\textbf{Model selection.} In this paper, we monitor the layer-wise neuron behaviors of a third-party DNN-based SR system. Speaker recognition systems aim at determining the identity of speakers by learning the acoustic features mostly with DNN-based models. In this work, we exploit the DNN-based SR system to serve as a third-party model for capturing the layer-wise neuron behaviors by leveraging its power in representing speech in a layer-wise manner.
% We can merely obtain the input and output pair like given a text and output a clip of synthetic voice. since their multi-layer design learn for the subtle differences between inputs.
% since we cannot obtain their parameters. Therefore, we need a third-party model based on DNNs to capture the neuron behaviors of input voices.

\textbf{Layer selection.} We select the layers that learn and preserve valuable representation information of inputs, such as convolutional and fully-connected layers in typical convolutional neural networks (CNNs). Here, other layers like pooling without learning substantial representation information can be seen as redundant layers. It might be interesting to explore layers that specifically learn the differences between real and fake voices in future work.

\textbf{Neuron coverage criteria.} We introduce two different neuron coverage criteria to figure out the threshold $\delta$ for determining the activated neurons. Then, the determined activated neurons in each selected layer are applied to represent the layer-wise behaviors of voices. Previous work \cite{pei2017deepxplore} uses a global threshold to determine if the neuron is activated or not, which is too coarse \cite{ma2018deepgauge}. Here, we specify each layer with a particular threshold. More details on calculating the threshold $\delta$ are presented in the following subsection.

% Specifically, we consider neurons in the layer-level by specifying the threshold for each layer rather than the whole model in previous work \cite{pei2017deepxplore} where a global threshold is defined to determine the neuron is activated or not, which is too coarse \cite{ma2018deepgauge}.

% which are applied for capturing the layer-wise neuron behaviors in the DNN model.

% ------------------------------
\begin{figure}[t]
\centering
\subfigure[MFCC]{
\includegraphics[width=0.28\columnwidth]{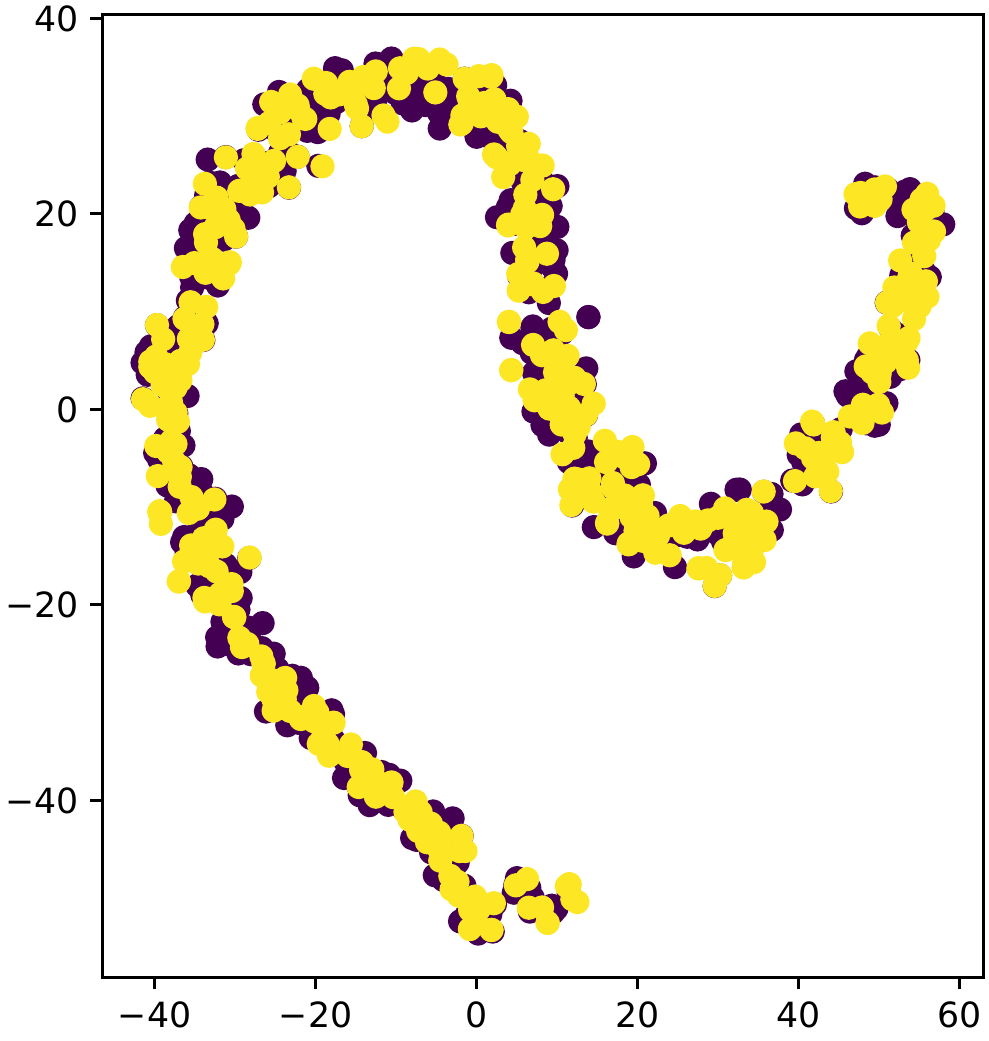}

}
\quad
\subfigure[Raw Neurons]{
\includegraphics[width=0.28\columnwidth]{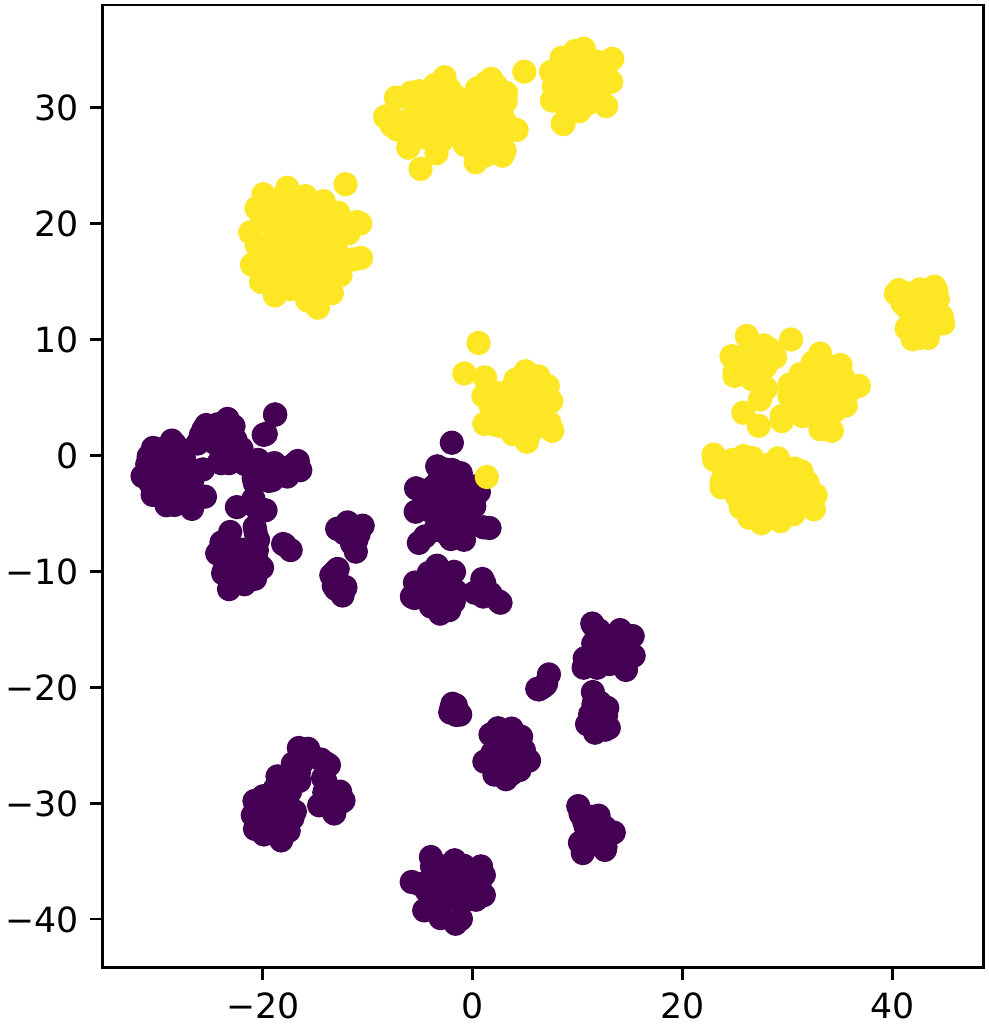}
}
\quad
\subfigure[Activated Neurons]{
\includegraphics[width=0.28\columnwidth]{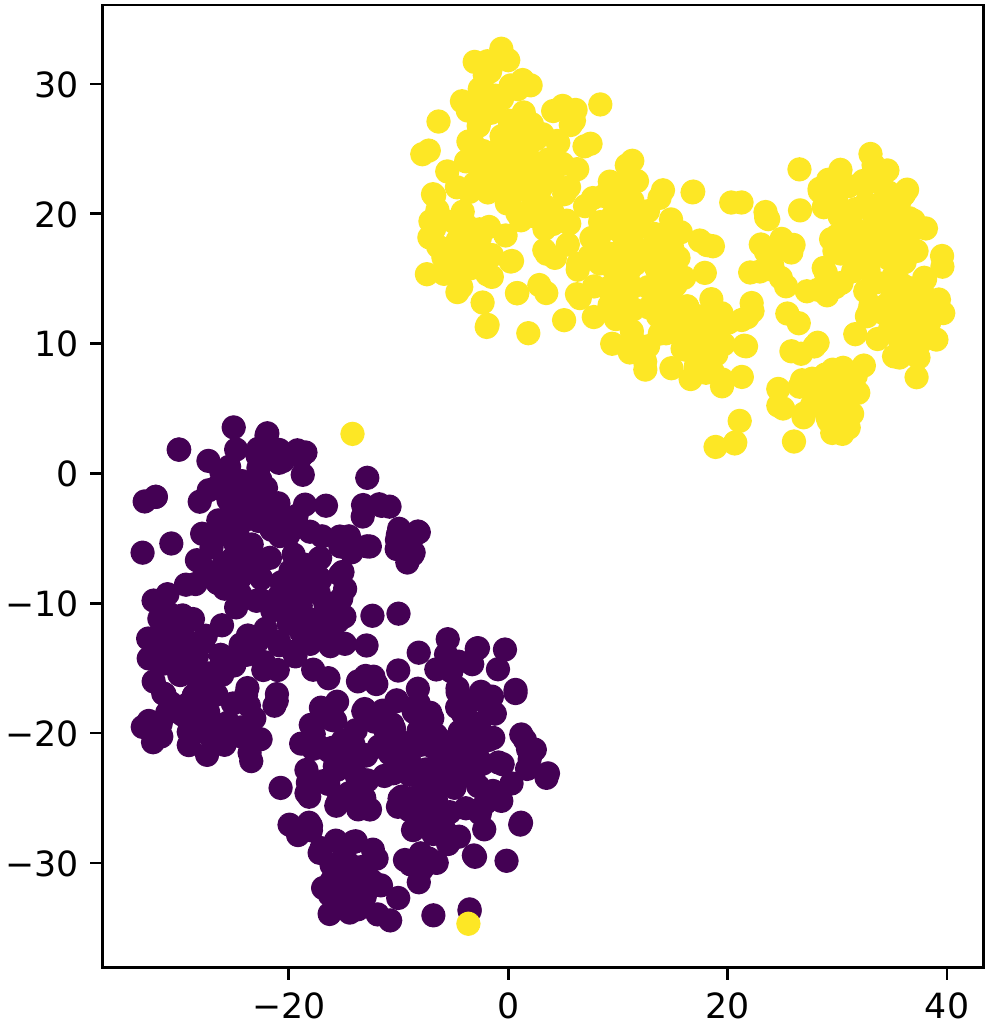}
}
\vspace{-10pt}
\caption{Visualization of three different features in representing real and fake voices. From the left to right, features are represented by MFCC, raw layer-wise neuron behaviors, and activated neuron behaviors with designed neuron coverage criteria, respectively. Here, we select the neuron coverage criteria TKAN as a presentation example.}
\vspace{-13pt}
\label{Figure:visual_neuron}
\end{figure}
% ------------------------------

\subsection{Neuron Coverage Criteria Design}

In this paper, we introduce two different neuron coverage criteria for determining the activated neurons to capture layer-wise neuron behaviors. The first one counts the number of activated neurons in each layer, called average count neuron (ACN). The other one selects neurons having top $k$ values in each layer, named Top-$k$ activated neuron (TKAN).

\textbf{ACN.} Motivated by the weakness of the global threshold defined in previous DNN testing studies, we specify each layer $l$ with a particular threshold $\delta_l$ that is calculated from the training dataset. The threshold $\delta_l$ is an average value of all the neuron output values in the layer $l$ of all inputs in the training dataset. We calculate the threshold $\delta_l$ with the following formula:
\begin{align} \label{eq:1}
\delta_l=\frac{\sum_{x \in X, i \in I}^{}\varphi(x,i;\theta)}{|I| \cdot |X|}
\end{align}
where $x$ is the neuron in $l$-th layer, $X$ is the set of neurons in layer $l$, $i$ is an input in the training dataset $I$, $\varphi$ is the activation function for calculating the neuron output value of input $i$ with trained parameter $\theta$, $|X|$ and $|I|$ represent the number of neurons in layer $l$ and the number of inputs in training dataset $I$, respectively. Here, we define the ACN as follows:
\begin{align} \label{eq:2}
\mathrm{\emph{ACN}}(l, i)=|\{{x|\forall x\in l, \varphi(x,i;\theta)>\delta_l}\}|
\end{align}
where $i$ represents the input, $x$ is the neuron in layer $l$, $\varphi$ is an activation function for computing the neuron output value, and $\delta_l$ is the threshold of the $l$-th layer calculated by formula (\ref{eq:1}).

\textbf{TKAN.} Instead of learning a threshold from training datasets to determine whether a neuron is activated or not, we explore another neuron coverage criterion by simply selecting neurons whose output value is ranked as top $k$ in its layer. Here, we conjecture that neurons with large output value are critical neurons that have high influences in representing inputs for a DNN model. We define the TKAN as follows:
\begin{align} \label{eq:3}
\mathrm{\emph{TKAN}}(l, i)= \{\mathop{\arg\max}_{k} (\varphi(x,i;\theta), k):x \in X \}
\end{align}
where the function $\mathop{\arg\max}$ returns $k$ neuron output values calculated with $\varphi$. Here, the $k$ is applied for all the layers in the model.

% We monitor the layer-wise neurons and determine the activated neurons with the designed neuron coverage criteria. Here,
Figure \ref{Figure:visual_neuron} adopts t-distributed stochastic neighbor embedding (T-SNE), an algorithm for high-dimensional data visualization, to visualize the effectiveness of neuron behaviors in hunting the differences between real and fake voices compared with Mel-scale frequency cepstral coefficients (MFCC), a popular feature in speech analysis. From L-R, voices are represented with MFCC, raw layer-wise neuron behaviors, and activated neurons with designed neuron coverage criteria, respectively. We can easily find that compared with MFCC, raw layer-wise neurons can capture the differences between real and fake in a coarse manners, where the voices are separated into several relatively independent clusters. Furthermore, the subtle differences between real and fake voices can be easily distinguished by applying our designed neuron coverage criteria, where real and fake are separated into two independent clusters.

\subsection{Fake Voice Detection}
We train a binary-classifier with a shallow neural network to predict whether a clip of voice is human speech or AI-synthesized fake voice. The inputs of our binary-classifier are the vectorized captured layer-wise neuron behaviors rather than the raw input of voices, which are better for a simple classifier to learn the differences between real and fake voices. Additionally, the neuron behavior inputs are insensitive to manipulations on voices, thus are robust against various manipulations, such as voice conversion and additive real-world noises.

Algorithm~1 describes our basic ideas of capturing layer-wise neurons behaviors for discerning real and fake voices. We train two supervised binary-classifiers with the same architecture based on the two different strategies, namely ACN and TKAN. In predicting an input, we first obtain the layer-wise neuron behaviors with ACN and TKAN, respectively. Then, the neuron behaviors are formed as input features into the binary-classifier for prediction. For ACN, the number of activated neurons in each layer is formed as a feature vector. For TKAN, the raw value of neuron output, which ranked the top $k$ in its layer is formed as a feature vector. Finally, the classifier predicts the voice based on the classifier’s final output score.

\section{Experimental Setting and Implementation}\label{sec:settings}

% In this section, we introduce the basic experimental settings, including datasets adopted for evaluation, baseline for comparison with prior work, and evaluation metrics. Additionally, we also present the implementation details.

\subsection{Dataset}

In our experiments, fake voices are collected from three different datasets including TTS and VC synthesized with various techniques. To ensure its diversity in languages and genders, English and Mandarin Chinese languages are spoken by males and females containing different accents. The first dataset is a public dataset, called \textbf{FoR}, created by APTLY lab \cite{lab} with the latest open-sourced tools and commercial speech synthesis products (\eg, Amazon AWS Polly, Google Cloud TTS, and Microsoft Azure TTS). The real voices in FoR are collected from open-sourced speech datasets and free available videos on internet like TED talks and YouTube videos, which cover a good variety of genders, speaker ages, and accents, \etc{} All the fake voices are synthesized with latest deep learning-based techniques, which own high qualities. However, the dataset FoR only contains the first type TTS fake voices that are synthesized by given texts.

% \lei{this sentence needs to rewrite}Unlike the public dataset released by Google \cite{Todisco2019} that adopts some outdated techniques for voice synthesis rather than state-of-the-art (SOTA) commercial products trained with massive powerful GPU resources.

Therefore, we build the second dataset, a VC fake voice dataset. The dataset is built by ourselves with an open-sourced tool sprocket \cite{kobayashi2018sprocket}, which allows to clone the source speaker's identity into the target speaker. Sprocket also served as a baseline system in voice conversion challenge 2018 (VCC18) \cite{lorenzo2018voice}. Here, real voices are collected from voice conversion challenge 2016 (VCC16) \cite{toda2016voice} and VCC18. The second dataset is called \textbf{Sprocket-VC}.
% while persevering linguistic content

\begin{SCfigure}[]
\centering
\includegraphics[width=0.45\columnwidth]{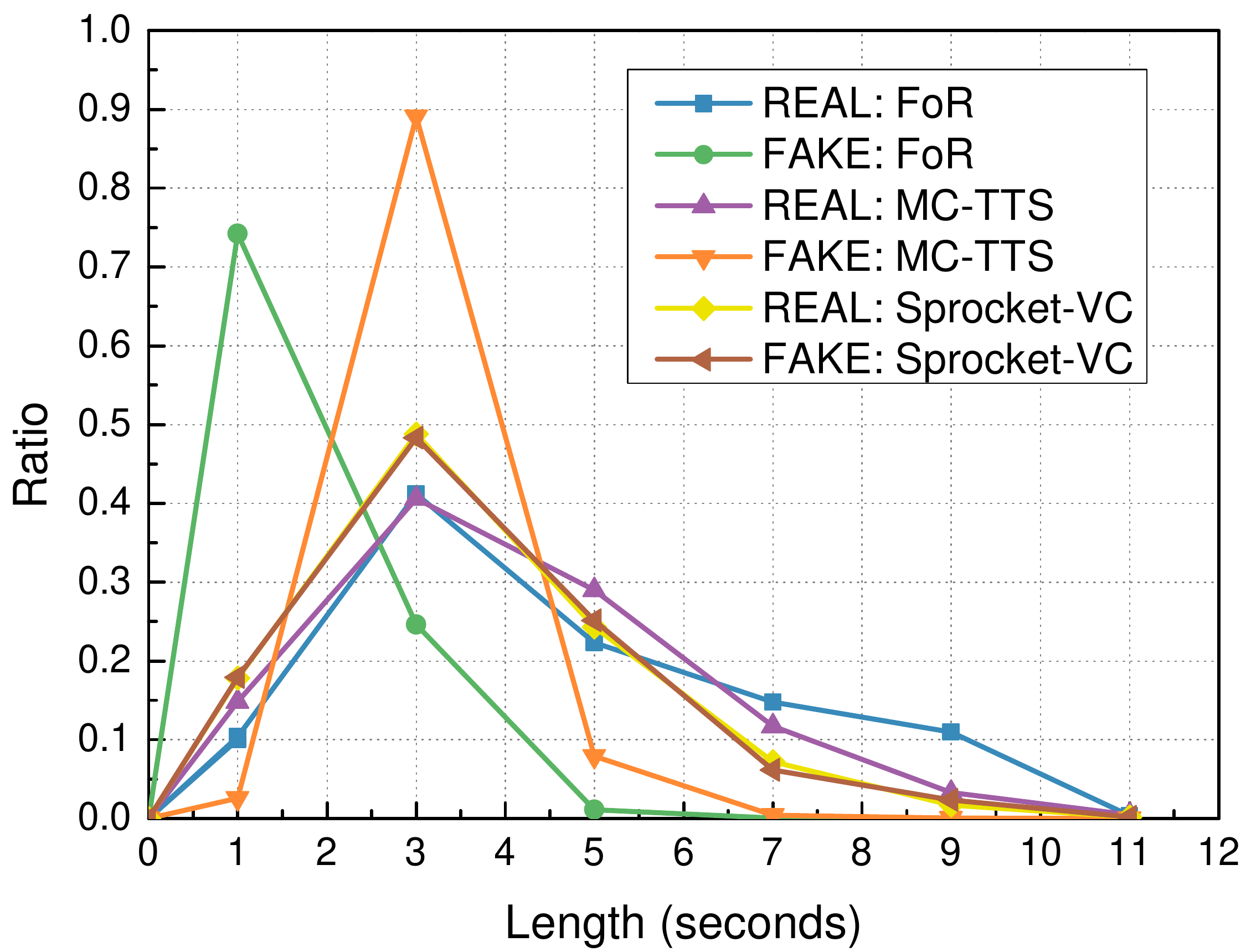}
\caption{Real and fake voices length distribution in the three datasets. Y-axis indicates the ratio of voices that lies in the length range with an offset $\pm$1.}
\label{Figure:audio_length_dis}
% \vspace{-16pt}
\end{SCfigure}

However, fake voices in the first and second datasets are all spoken in English language, thus we build the third dataset, where fake voices are all spoken in Mandarin Chinese for evaluating the capabilities of our approach in tackling different languages. We adopt the Baidu speech synthesis system \cite{baidu_tts} that achieves the best performance in Chinese language synthesis. We give a series of ancient poetry \cite{poetry} as input texts to produce numerous fake voices. The third dataset is called \textbf{MC-TTS}. More details of the three datasets are summarized in Table \ref{Table:data_collection}. We also present the length distribution of voices in the three datasets in Figure \ref{Figure:audio_length_dis}.

\begin{table}[]
\scriptsize
\centering
\caption{Statistics of the three datasets for evaluating the effectiveness and robustness of DeepSonar. Column \emph{Language} indicates the language spoken in the voice samples. Column \emph{Real Voice Collection} means the sources of real voices collected in the dataset. All the real and fake voices in FoR are collected from the second version \emph{for-norm} in the original dataset where three different versions are included. Column \emph{Model} represents the number of techniques for synthesizing voices. Last two columns \emph{Real(\#)} and \emph{Fake(\#)} denote the number of real and fake voices in each dataset.}
\vspace{-10pt}
\setlength{\tabcolsep}{3.2pt}
\begin{tabular}{c|c|c|c|c|c|c}
\toprule
\textbf{Dataset} & \textbf{Type} & \textbf{Language} & \textbf{Real Voice Collection} & \textbf{Model} & \textbf{Real(\#)} & \textbf{Fake(\#)} \\ \midrule
FoR     &   TTS &   English&    multi-sources & 7  & 26,941  & 26,927 \\ \midrule
MC-TTS&      TTS&    Chinese&    lecture\_tts \cite{lrcd}& unknown & 6,000& 6,026\\ \midrule
Sprocket-VC &   VC&     English&    VCC16\&VCC18 &   1   &  3,132 & 3,456\\
\bottomrule
\end{tabular}
\label{Table:data_collection}
\vspace{5pt}
\includegraphics[width=0.82\columnwidth]{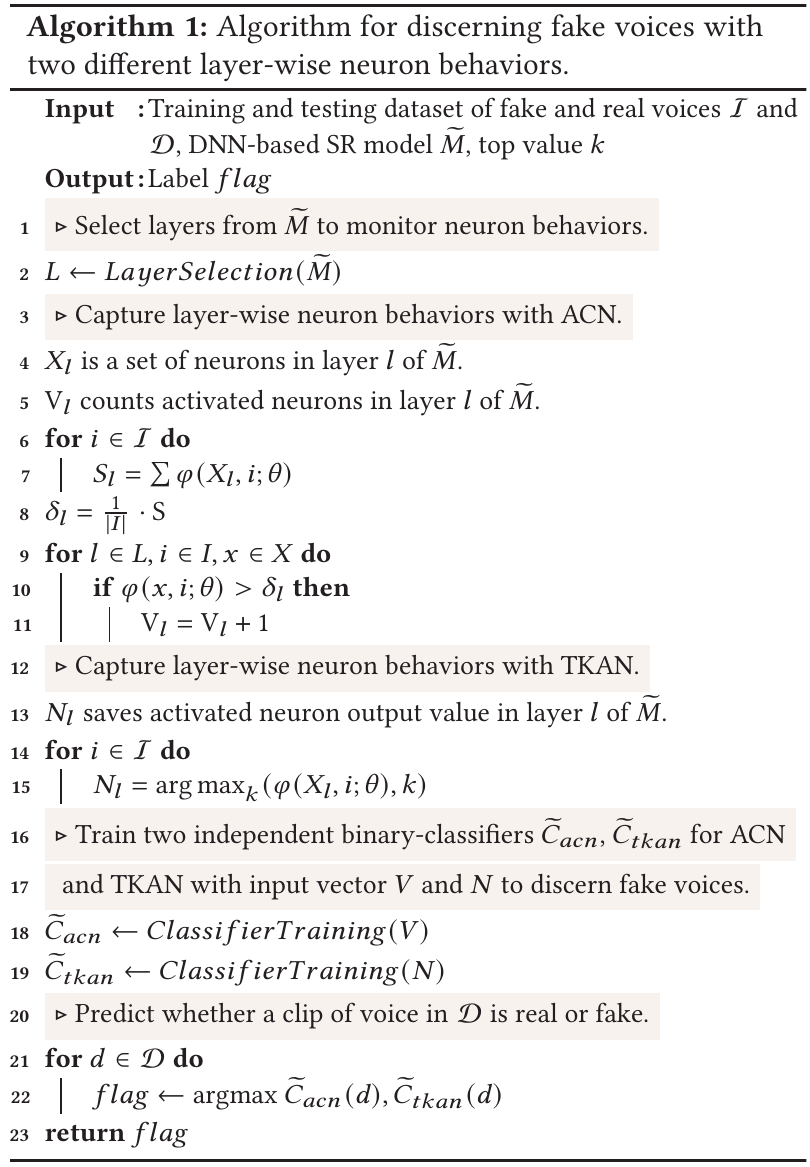}
\vspace{-20pt}
\end{table}
%------------------------
%------------------------
% \begin{figure}
% \centering
% \includegraphics[width=0.5\columnwidth]{image/length_dis_new.pdf}
% \caption{Real and fake voices length distribution in the three datasets. Y-axis indicates the ratio of voices that lies in the length range with an offset $\pm$1.}
% \label{Figure:audio_length_dis}
% % \vspace{-15pt}
% \end{figure}

%------------------------

\subsection{Baseline}

In evaluation, we mainly compared our work with a prior work leveraging bispectral artifacts on fake voices to differentiate human speech and AI-synthesized fake voices \cite{albadawy2019detecting}. To the best of our knowledge, this is a SOTA work focused on AI-synthesized fake voice detection. We implemented this work with open-sourced available repositories in GitHub \cite{bispectral_audio}. The details of the baseline are introduced in Section \ref{sec:dr}.

% the other one is an open-sourced project for detecting audio DeepFakes, which utilizes mel-frequency spectrograms as input features to train a DNN model by performing temporal convolutions on spectrograms to discern fake voices \cite{temporal_audio}.

\subsection{Evaluation Metrics}
For a comprehensive evaluation of DeepSonar, we adopt seven different metrics to evaluate the capabilities of DeepSonar in fighting against TTS and VC in the three datasets.

Specifically, we use accuracy, AUC (area under curve) of ROC (receiver operating characteristics), F1-score, and AP (average precision) to evaluate whether DeepSonar achieves a higher detection rate. We use FPR (false positive rate), FNR (false negative rate), and EER (equal error rate) to get the false alarm rate of DeepSonar in prediction. These seven metrics are widely served as metrics in evaluating the performance of classifiers.
% binary-classifier in various tasks, such as malware detection.

\subsection{Implementation}
We design a shallow neural network with five fully-connected layers as our binary-classifier for discerning fakes. The optimizer is SGD with momentum 0.9 and the starting learning rate is 0.0001, with a decay of 1e-6. The loss function is binary cross-entropy.

In monitoring neuron behaviors, we employ a speaker recognition deep network that adopts a `thin-ResNet' as its backend architecture \cite{Xie19a} and select the convolutional and fully-connected layers to capture the layer-wise neuron behaviors as input features. Our approach is generic to any speech representation system, which could be easily extended to other systems that have the capability to learn speech representations layer-by-layer. For TKAN, we empirically set $k$ to $5$ with a consideration of the number of selected layers and training samples.
% If we have large training data, it is fine to set a large $k$.
% It is interesting to explore how to define $k$ in our future work.
% In evaluating the robustness of DeepSonar against manipulation attacks, we select more than $15$ different manipulations to get a comprehensive evaluation.
%
In evaluating the robustness of DeepSonar against manipulation attacks, we select more than $15$ different manipulations to achieve a comprehensive evaluation. We hope that these $15$ different representative voice manipulations could also serve as a robustness evaluation benchmark for future research. Table \ref{Table:manipulation_attacks} shows the $15$ different manipulations, which are classified as voice conversions by changing voice signals and real-world noises by adding environmental noises. The real-world noise samples are collected from a public dataset ESC-50 that includes lots of environmental audio recordings \cite{piczak2015dataset}.
% more than $2,000$
%------------------------
\begin{table}[t]
\scriptsize
\centering
\caption{Voice conversions and additive real-world noises in manipulation attacks. Voice conversion includes three common transformations when publishing audios. Additive real-world noises are classified into indoor and outdoor environmental sounds. The selected $12$ real-world noises from ESC-50 are representative environmental sounds in real scenarios.}
\vspace{-10pt}
% which could be easily adopted by attackers to evade detection intentionally.
\setlength{\tabcolsep}{3.5pt}
\begin{tabular}{c|c|c}
\toprule
\textbf{Manipulation Attacks} & \multicolumn{2}{c}{\textbf{Sound Classes}} \\ \midrule
Voice Conversions     &   \multicolumn{2}{c}{1) resampling, 2) speed, 3) pitch}  \\ \midrule
\multirow{4}{*}{Real-world Noises}
    & \multirow{2}{*}{Indoors} & \multicolumn{1}{l}{1) breathing, 2) footsteps, 3) laughing}    \\ \cline{3-3}
    &   & \multicolumn{1}{l}{4) mouse-click, 5) keyboard-type, 6) clock-tick}    \\ \cline{2-3}
    & \multirow{2}{*}{Outdoors} & \multicolumn{1}{l}{1) engine, 2) train, 3) fireworks}    \\ \cline{3-3}
    & & \multicolumn{1}{l}{4) rain, 5) wind, 6) thunderstorm}    \\
\bottomrule
\end{tabular}
\label{Table:manipulation_attacks}
\vspace{-20pt}
\end{table}
%------------------------

% All our experiments are conducted on a server running Ubuntu 16.04 system on a total of 40 cores 2.20GHz Xeon CPUs with 500GB RAM and four NVIDIA Tesla V100 GPUs with 36GB memory for each.
% 2.20GHz

\section{Experimental Results}\label{sec:exp_results}

% In this section, we conduct experiments in three datasets to comprehensively evaluate the effectiveness of DeepSonar in discerning AI-synthesized fake voices and its robustness against two typical manipulation attacks, voice conversions and additive real-world noises. Thus, our evaluation aims to answer the following two research questions.

Our evaluation aims to answer the following research questions.

\begin{itemize}[leftmargin=*]
\item \textbf{RQ1}: What is the performance of DeepSonar in discerning two types of fake voices (TTS and VC) synthesized with various techniques and tackling different languages?
\item \textbf{RQ2}: Whether DeepSonar is robust against voice manipulation attacks including voice conversions and additive real-world noises at various magnitudes?
\end{itemize}

\subsection {Detection Results (RQ1)} \label{sec:dr}

In this section, we mainly answer the first research question, \ie, whether our approach DeepSonar can effectively discern real and fake voices and tackle different languages. Our experiments are conducted on the three different datasets (see Table \ref{Table:data_collection}). Each dataset is splitted into three parts, \eg{}, 60\%, 20\%, 20\% as training, validation and testing, respectively. Specifically, we also compared our work with previous work using bispectral artifacts (served as a baseline) and report the detection rate and false alarm rate using seven different metrics.

\textbf{Effectiveness of DeepSonar.} Table \ref{Table:acn_tkan} summarizes the experimental results of DeepSonar using two different neuron coverage criteria for determining activated neurons. DeepSonar gives an average accuracy >\textbf{98.1\%} and an EER <\textbf{2\%} on the three datasets and demonstrates the effectiveness in discerning the two typical fake voices in both English and Chinese languages. In the first dataset FoR where voices are synthesized with commercial products and more challenging than the other two datasets, DeepSonar obtains an accuracy >99\% when employing TKAN, but it reaches an accuracy <90\% when adopting ACN. This result illustrates that using TKAN can be more powerful than ACN in tackling voices synthesized with various commercial-level synthetic techniques. Thus, we mainly compare our approach using TKAN with the baseline.

\textbf{Compared with baseline.} Table \ref{Table:effectiveness} summarizes the results compared with the baseline. Both the baseline and our proposed DeepSonar are trained and tested on the same datasets. Experimental results show that the average performance of DeepSonar using TKAN significantly outperforms the baseline on the three datasets. The baseline is a SOTA work using bispectral artifacts in fake voices to differentiate real and fake voices \cite{albadawy2019detecting}. They found that higher-order spectral correlations rarely exist in real human speech while they are common in AI-synthesized fake voices. In their experiments, a simple classifier with SVM is adopted to identify the bispectral artifacts for differentiating real and fake voices. Different from this work investigating the artifacts introduced in synthesis, we leverage the power of layer-wise neuron behaviors for representing inputs, which provides cleaner signals than raw voice inputs (\eg{}, bispectral artifacts in voices) for simple binary-classifier in hunting the differences between real and fake voices.

According to the experimental results in Table \ref{Table:acn_tkan} and Table \ref{Table:effectiveness}, detecting clean AI-synthesized fake voices without any degradation is a relatively easy task by DeepSonar. Unfortunately, voice manipulations like voices resampling, adding real-world noises are common in real applications, thus evading manipulation attacks is important for detectors deployed in the wild. In the next subsection, we mainly discuss the robustness of our approach in tackling manipulation attacks at various magnitudes.

%------------------------
\begin{table}[]
\scriptsize
\centering
\caption{Performance of DeepSonar using two different neuron behaviors (\eg{}, ACN and TKAN) in discerning real and fake voices. The last row \emph{DeepSonar} represents an average results on the three datasets. $\uparrow$ means the larger value the better, while $\downarrow$ indicates the smaller value the better.}
\vspace{-10pt}
\setlength{\tabcolsep}{3.0pt}
\begin{tabular}{c|c|c|c|c|c||c|c|c}
\toprule
\textbf{Datasets} & \textbf{Methods} & \textbf{Acc. $\uparrow$} & \textbf{AUC $\uparrow$} & \textbf{F1 $\uparrow$} & \textbf{AP $\uparrow$} & \textbf{FPR $\downarrow$} & \textbf{FNR $\downarrow$} & \textbf{EER $\downarrow$}\\ \midrule

\multirow{2}{*}{FoR} & \textbf{ACN} & 0.8927  & 0.8930  & 0.8939 & 0.8604 & 0.1193 & 0.0946 & 0.1164 \\ \cline{2-9}
                      & \textbf{TKAN} & 0.9998  & 0.9998  & 0.9998 & 0.9997 & 0.0002 & 0.0002 & 0.0002 \\ \midrule
\multirow{2}{*}{\makecell[c]{Sprocket-VC}} & \textbf{ ACN} & 0.9989  & 0.9989  & 0.9989 & 0.9989 & 0.002 & 0.0 & 0.002 \\ \cline{2-9}
                             & \textbf{TKAN}& 1.0  & 1.0  & 1.0 & 1.0 & 0.0 & 0.0 & 0.0 \\ \midrule
\multirow{2}{*}{MC-TTS} & \textbf{ACN} & 0.9975  & 0.9975  & 0.9975 & 0.9975 & 0.005 & 0.0 & 0.005 \\ \cline{2-9}
                         & \textbf{TKAN}& 1.0  & 1.0  & 1.0 & 1.0 & 0.0 & 0.0 & 0.0 \\ \midrule
\multicolumn{2}{c|}{\textbf{DeepSonar}}& 0.981  & 0.982  & 0.982 & 0.976 & 0.021 & 0.016 & 0.021 \\
\bottomrule
\end{tabular}
\label{Table:acn_tkan}
% \vspace{-10pt}
\end{table}
%------------------------
%------------------------
\begin{table}[]
\scriptsize
\centering
\caption{Performance of DeepSonar and the baseline using bispectral artifacts based from Farid \etal \cite{albadawy2019detecting}, on three datasets in discerning AI-synthesized fake voices. DeepSonar utilizes TKAN to monitor neuron behaviors. Average result denotes an average performance of the three approaches on the three different datasets measured by seven metrics.  $\uparrow$ means the larger value the better, while $\downarrow$ indicates the smaller value the better.}
\vspace{-10pt}
\setlength{\tabcolsep}{3.0pt}
\begin{tabular}{c|c|c|c|c|c||c|c|c}
\toprule
\textbf{Datasets} & \textbf{Methods} & \textbf{Acc. $\uparrow$} & \textbf{AUC $\uparrow$} & \textbf{F1 $\uparrow$} & \textbf{AP $\uparrow$} & \textbf{FPR $\downarrow$} & \textbf{FNR $\downarrow$} & \textbf{EER $\downarrow$}\\ \midrule

\multirow{2}{*}{FoR}  & Farid \etal \cite{albadawy2019detecting} & 0.713  & 0.746  & 0.757 & 0.821 & 0.345 & 0.163 & 0.292 \\ \cline{2-9}
                    %   & Dessa \cite{temporal_audio} & 0.986  &  0.987 & 0.987 & 0.984 & 0.021 & 0.006 & 0.020 \\ \cline{2-9}
                    %   & \textbf{DeepSonar (ACN)} & 0.8927  & 0.8930  & 0.8939 & 0.8604 & 0.1193 & 0.0946 & 0.1164 \\ \cline{2-9}
                      & \textbf{DeepSonar} & 0.9998  & 0.9998  & 0.9998 & 0.9997 & 0.0002 & 0.0002 & 0.0002 \\ \midrule
\multirow{2}{*}{\makecell[c]{Sprocket \\ VC}} & Farid \etal \cite{albadawy2019detecting} & 0.652  & 0.658  & 0.681 & 0.687 & 0.371 & 0.314 & 0.351 \\ \cline{2-9}
                            %  & Dessa \cite{temporal_audio} & 1.0  & 1.0  & 1.0 & 1.0 & 0.0 & 0.0 & 0.0 \\ \cline{2-9}
                            %  & \textbf{DeepSonar (ACN)} & 0.9989  & 0.9989  & 0.9989 & 0.9989 & 0.002 & 0.0 & 0.002 \\ \cline{2-9}
                             & \textbf{DeepSonar}& 1.0  & 1.0  & 1.0 & 1.0 & 0.0 & 0.0 & 0.0 \\ \midrule
\multirow{2}{*}{MC-TTS}  & Farid \etal \cite{albadawy2019detecting} & 0.626  & 0.693  & 0.711 & 0.869 & 0.421 & 0.193 & 0.343 \\ \cline{2-9}
                        %  & Dessa \cite{temporal_audio} & 0.992  & 0.992  & 0.992 & 0.984 & 0.0 & 0.016 & 0.016 \\ \cline{2-9}
                        %  & \textbf{DeepSonar (ACN)} & 0.9975  & 0.9975  & 0.9975 & 0.9975 & 0.005 & 0.0 & 0.005 \\ \cline{2-9}
                         & \textbf{DeepSonar}& 1.0  & 1.0  & 1.0 & 1.0 & 0.0 & 0.0 & 0.0 \\ \midrule
\multirow{2}{*}{\textit{\makecell[c]{Average \\ Result}}}  & Farid \etal \cite{albadawy2019detecting} & 0.664  & 0.699  & 0.716 & 0.792 & 0.379 & 0.223 & 0.329 \\ \cline{2-9}
                            % & Dessa \cite{temporal_audio} & 0.993  & 0.993  & 0.993 & 0.989 & 0.007 & 0.007 & 0.012 \\ \cline{2-9}
                            % & \textbf{DeepSonar (ACN)}& 0.963  & 0.963  & 0.963 & 0.952 & 0.042 & 0.032 & 0.041\\ \cline{2-9}
                            & \textbf{DeepSonar}& 1.0  & 1.0  & 1.0 & 1.0 & 0.0 & 0.0 & 0.0 \\
                            % & \textbf{DeepSonar}& 0.981  & 0.982  & 0.982 & 0.976 & 0.021 & 0.016 & 0.021 \\
\bottomrule
\end{tabular}
\label{Table:effectiveness}
% \vspace{-15pt}
\end{table}

\subsection{Evaluation on Robustness (RQ2)}

The biggest difference between AI-synthesized fake images and fake voices lies in that manipulations like voice conversion and additive real-world noises can be easily camouflaged as regular operations. In this section, we evaluate the robustness of DeepSonar in tackling voice conversion and additive real-world noises at various magnitudes to investigate the second research question.
%------------------------
\begin{figure}[t]
\centering
\includegraphics[width=0.49\columnwidth]{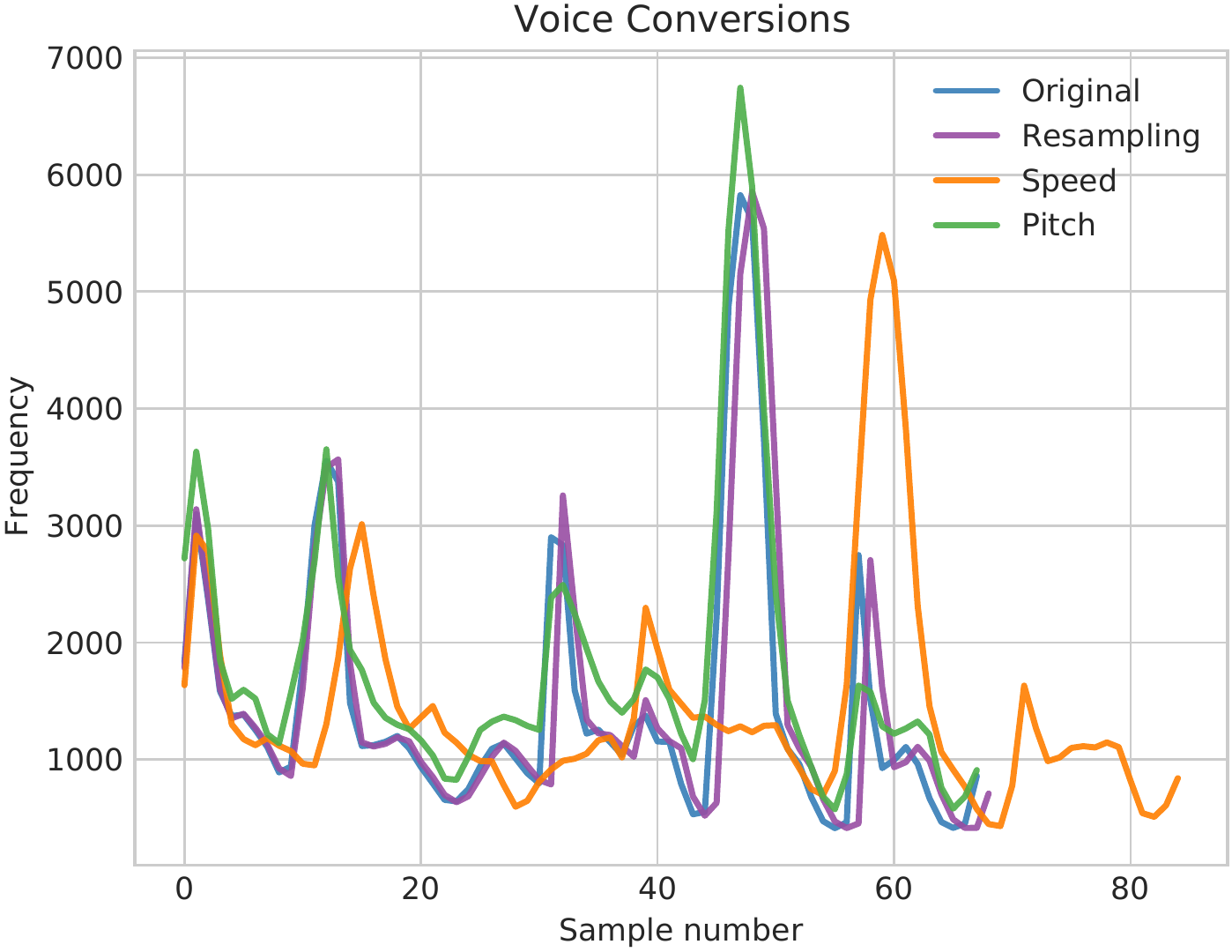}
\includegraphics[width=0.49\columnwidth]{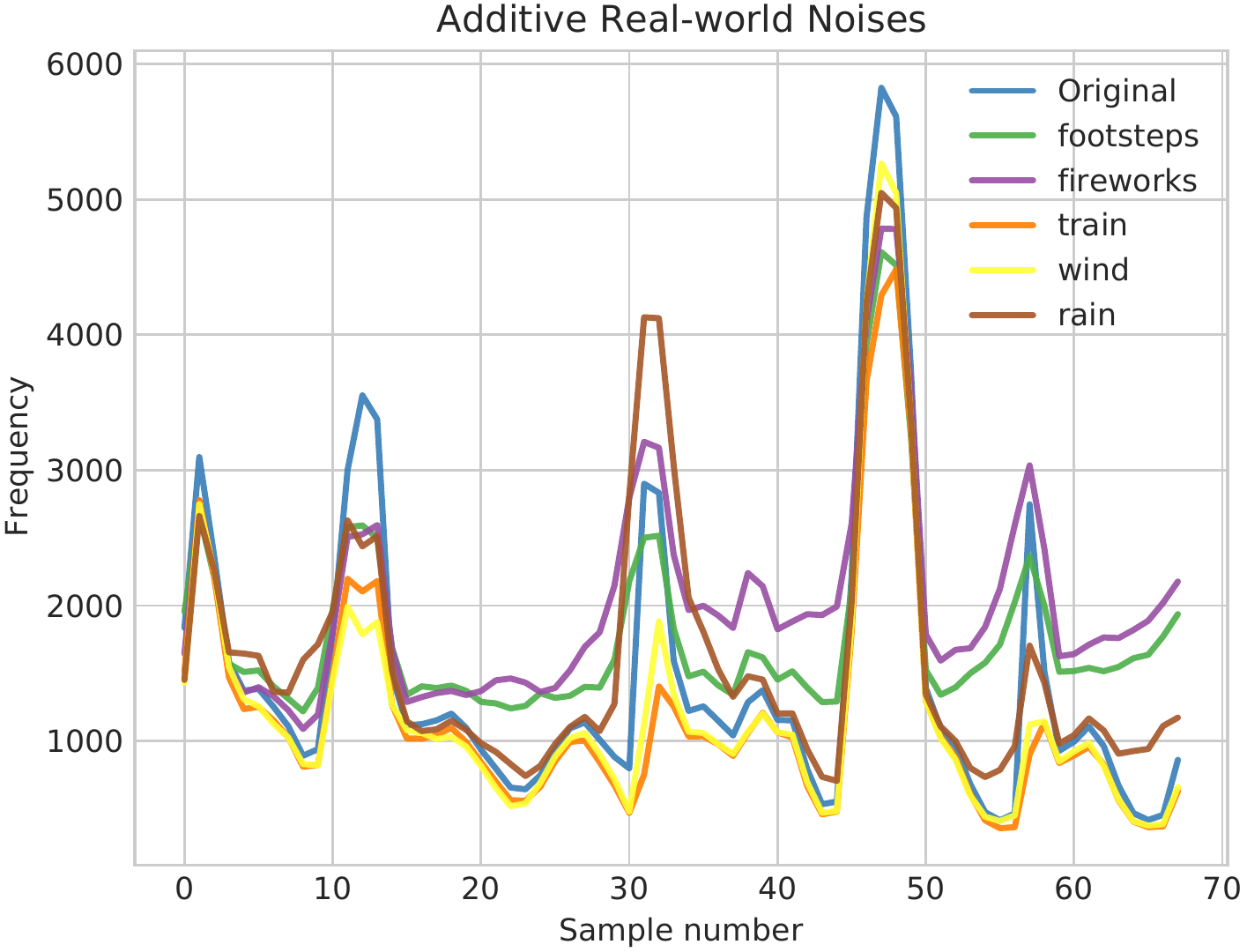}
\vspace{-8pt}
\caption{Signals of voices manipulated by voice conversion and additive real-world noises. In voice conversions, voice upsampling by adding 400, speed ratio set to 0.8 times, pitch-shifted by 4 steps. In additive real-world noises, we present voice signals by adding four real-world noises including indoors (footsteps) and outdoors (fireworks, train, wind, thunderstorm) noises (SNR=35). The original clip of synthesized fake voice is from the FoR dataset saying "\textit{Do you feel like eating something}".}
\label{Figure:noises}
\vspace{-12pt}
\end{figure}
%------------------------

% In experiments, we mainly focus on two different manipulation attacks, voice conversion without modifying linguistic content, and additive real-world noises adding indoor or outdoor environmental noises. More details on the specific manipulations refer to Table \ref{Table:manipulation_attacks}.
% Voice conversion manipulates the voices without modifying linguistic content. Additive real-world noise adds indoor or outdoor environmental noises into voice involving linguistic content modification.
% For example, resampling the voice for publishing purpose, adding real-world noises since voices are recorded in noisy environments such as outdoors.

\textbf{Experimental settings.} In experiments, we select $1,000$ samples including $500$ real and $500$ fake voices from the testing dataset in FoR since they are synthesized with commercial products and more challenging for detection. We also employ TKAN for DeepSonar and compare it with the baseline like in effectiveness evaluation experiment. AUC is adopted for evaluation metrics as it is often used in the binary-classifier performance evaluation. Additionally, we use signal to noise ratio ($\mathrm{SNR}$) as metrics to evaluate the magnitude of real-world noises. The $\mathrm{SNR}$ is defined as
% The DNN-based baseline using spectrograms is adopted for comparison as it achieves competitive results in effectiveness evaluation than an artifact-based approach leveraging bispectral artifacts. Here, the layer-wise neuron behaviors are captured with TKAN since they are more robust than ACN by calculating thresholds from the training dataset to determine activated neurons. Additionally, it achieves competitive results on FoR than using ACN.
% \begin{align}
%     \mathrm{SNR} = 20 \log\Bigg(\frac{\mathrm{RMS}^2_{\mathrm{signal}}}{\mathrm{RMS}^2_{\mathrm{noise}}}\Bigg)
% \end{align}
$\mathrm{SNR} = 20 \log\Big(\frac{\mathrm{RMS}^2_{\mathrm{signal}}}{\mathrm{RMS}^2_{\mathrm{noise}}}\Big)$, where $\log(\cdot)$ is the logarithm of base $10$ and $\mathrm{RMS}$ is the root mean square.

By adding noises to voice data, we first need to obtain the $\mathrm{RMS}$ of the noises and voices, respectively. Then, we modify the noise by multiplying each element with a constant to change the $\mathrm{RMS}$, thus the desired $\mathrm{SNR}$ is achieved. In voice conversion, various voice manipulations are implemented with the APIs provided by libsora \cite{librosa}. Figure \ref{Figure:noises} presents a spectral centroid visualization of the two manipulation attacks, the left is voice conversion and the right is additive real-world noises. The two manipulations all have obvious modifications to the signals, which poses challenges to detectors.

\begin{figure*}[t!]
\centering
\subfigure[Voice Conversions]{
\includegraphics[width=0.6\columnwidth]{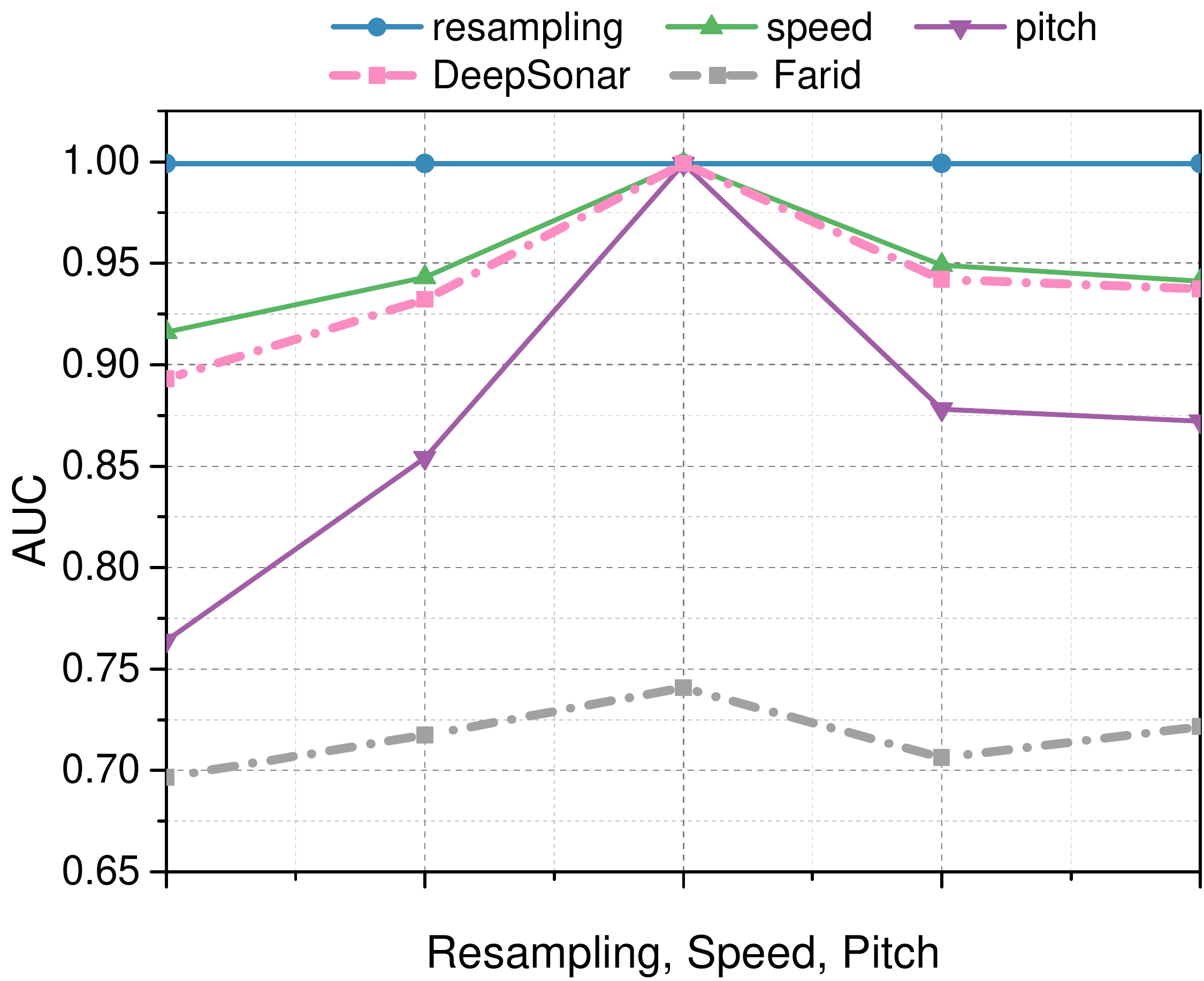}
}
\quad
\subfigure[Additive Real-world Noises (Indoors)]{
\includegraphics[width=0.6\columnwidth]{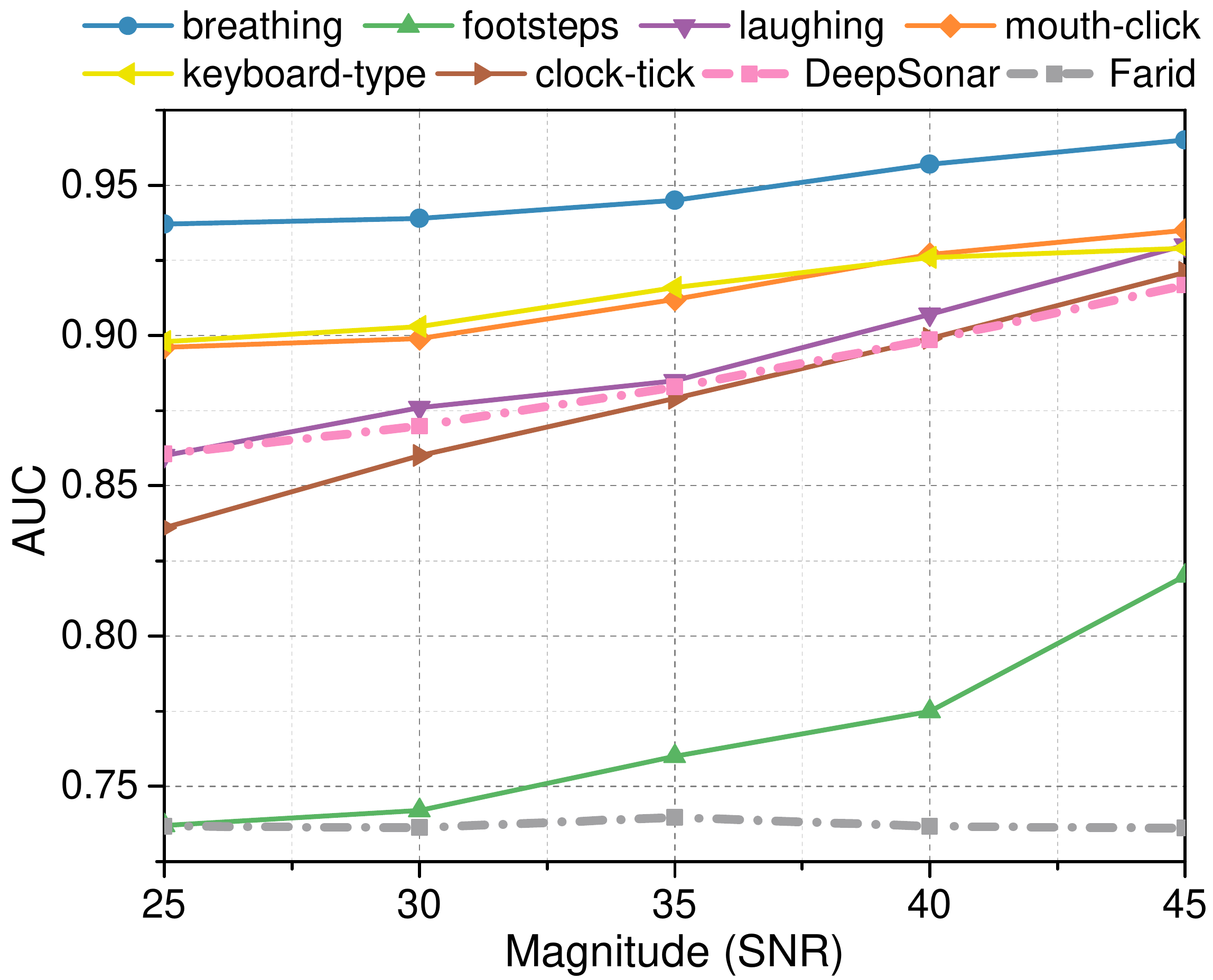}
}
\quad
\subfigure[Additive Real-world Noises (Outdoors)]{
\includegraphics[width=0.6\columnwidth]{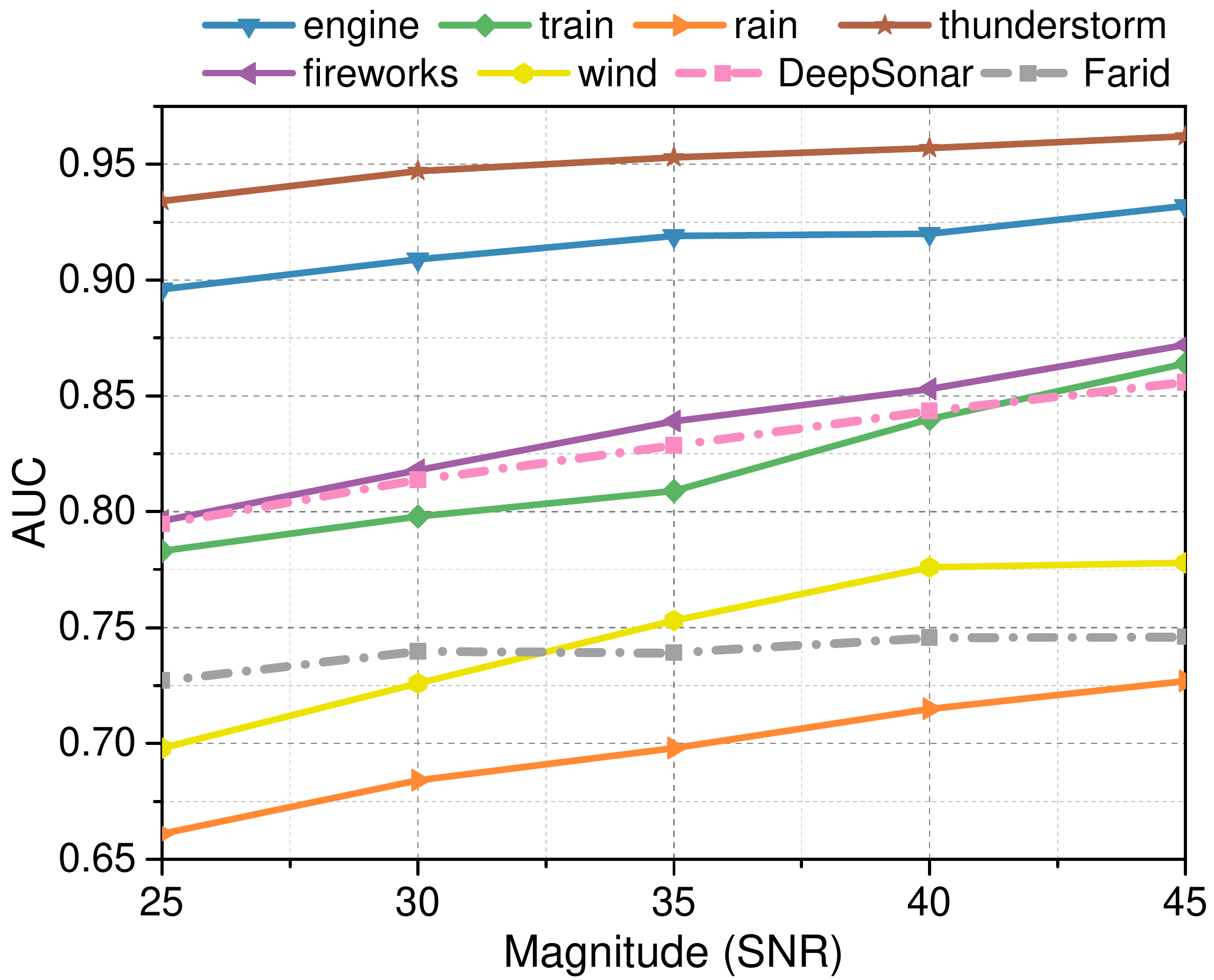}
}
\vspace{-10pt}
\caption{Robustness evaluation of DeepSonar against manipulation attacks at various magnitudes. In voice conversions (a), values in x-axis for resampling are \{-400, -200, 0, 200, 400\}, for speed are \{0.5, 0.8, 1.0, 1.2, 1.4\}, for pitch are \{-4, -2, 0, 2, 4\}. The dotted lines in the three subfigures represent an average performance of our approach DeepSonar and the baseline over different voice manipulations at various magnitudes. Large $\mathrm{SNR}$ means less noises added.}
\label{Figure:robustness}
\vspace{-10pt}
\end{figure*}

\textbf{Results on voice conversions.} Figure \ref{Figure:robustness}(a) shows the experimental results of DeepSonar in tackling three typical voice conversions. We could observe that DeepSonar is robust against resampling including upsampling and downsampling without any performance affected. The average performance is decreased less than 5\% and 15\% in stretching the voices and shifting pitches, respectively. Compared to the other two conversions (resampling and speed), DeepSonar seems to be a little susceptible to pitch shifting. The main reason is that voices with pitch-shifting have been broken and can hardly listen to the words in voices when the \emph{n\_steps} for changing the pitch of voices is larger than $2$. The settings for the three voice conversions are presented as follows.

In voice conversion, resampling indicates a time series of voice that is resampled from the original sample rate to the target sample rate, including upsampling and downsampling. Here, the target sample rate is set with an offset (\eg{}, $-400, 200, 0, 200, 400$) to the original sample rate, where offset $0$ servers as a baseline without resampling. Speed represents time-stretch an audio series by a fixed rate. The fixed-rate is set to $0.5, 0.8, 1.0, 1.2, 1.4$, where $1.0$ serves as a baseline. Pitch means we shift the pitch of a waveform by n\_steps semitones. Here, the n\_step is set to $-4, -2, 0, 2, 4$, where n\_step $0$ serves as a baseline that no pitch is shifted.
% -------------------------------------------------------
% -------------------------------------------------------

% Experimental results in Figure \ref{Figure:robustness} (a) show that DeepSonar is robustness against resampling including upsampling and downsampling without any performance affected. The average performance is decreased less than 5\% and 15\% in stretching the voices and shifting pitches, respectively. We listened the voices with pitch shifting and find that the voices has been broken and hardly listen the words they are speaking when the n\_step is larger than $2$.

% Experimental results in Figure \ref{Figure:robustness} (a) shows that DeepSonar is susceptible to pitch conversion compared to the other two conversion like resampling and speed.
% Figure \ref{Figure:robustness} presents us the robustness evaluation of DeepSonar against two manipulation attacks, voice conversion and additive real-world noises, in the first dataset FoR.

\textbf{Results on indoor-noises.} In additive real-world noises, voices are added with representative indoors and outdoors environmental noises. We use $\mathrm{SNR}$ to measure the magnitudes of added-noises. In Figure \ref{Figure:robustness}(b), DeepSonar performs well on the five indoor noises and the average performance decreased less than 10\% at the total five different magnitudes. However, the average performance is decreased by nearly 20\% at the five magnitudes when adding footstep noises. We listened to the added-footstep voices which have obviously mixed sizzle noises caused by the friction with floors. Figure \ref{Figure:noises} also visualizes the differences between original voices and added-footsteps voices.
% added with footsteps
% in all the noises without significant performances reduction except adding footstep noises where an average performance decreased nearly 20\% on all the five different magnitudes. The average performance on the six indoor environmental noises is decreased less than 12\% on all the five different magnitudes.

\textbf{Results on outdoor-noises.} In Figure \ref{Figure:robustness}(c), outdoor environmental noises can be roughly classified into three different categories based on the performance of DeepSonar. Engine and thunderstorm environmental noises are the first categories, where the average performance of DeepSonar decreased less than 7\% at the five different magnitudes. Fireworks and trains are the second categories, where the average performance of DeepSonar decreased less than 18\% at the five different magnitudes. Wind and rain are the third categories, where the average performance of DeepSonar decreased by nearly 25\% at the five different magnitudes. We find that environmental noises wind and rain also mixed with other voices like raindrops on the ground which is much noisy than other types of real-world environmental noises.

According to the experimental results in Figure \ref{Figure:robustness}, DeepSonar is also robust against voice conversions except voices are seriously damaged like shifting pitch with a big step. Additionally, DeepSonar performs well when the additive real-world noises are single voice without any mixture with other types of noises. In tackling mixed noises like wind, DeepSonar also holds a high detection performance at the magnitude measured by $\mathrm{SNR}$ larger than $35$.

\textbf{Compared with baseline.} The dotted lines in the three subfigures of Figure \ref{Figure:robustness} show the comparison results with the baseline by using bispectral artifacts to discern fake voices. To compare the performance of robustness with baseline, we use the average results of the two approaches over different types of voice manipulations at various magnitudes. For example, in Figure \ref{Figure:robustness}(b), each point in the dotted line is an average AUC score of the six additive indoor noises at the same magnitude. In Figure \ref{Figure:robustness}, the dotted line of DeepSonar is above the baseline, indicating that DeepSonar significantly outperforms the baseline in the two manipulation attacks.
% Experimental results in Figure \ref{Figure:robustness} show that DeepSonar significantly outperforms the baseline in all the two manipulation attacks.
% outperforms the baseline in the three different manipulation attacks. The result is an average results of two approaches over different voices manipulations at various magnitudes.

\subsection{Discussion}

DeepSonar achieves competitive results in terms of both effectiveness, and robustness against two manipulation attacks. However, DeepSonar also exhibits some limitations. First, in adversarial environments, adversaries could add an additional loss function by modeling the neuron behaviors to generate adversarial voices and evade detection. However, most learning-based approaches suffer this adversarial noise attack and an obvious trade-off between generating adversarial voices and evading detection exists. Secondly, real-world noises with a mixture of other types of noises at a high magnitude could decrease the performance of DeepSonar to some extent. Voice denoising will be a potential strategy for high-intensity mixed noises, which would be our future work to remove additional environmental noises. Especially, the voice denoising component is effective without obtaining any prior knowledge of the noises in the complex environments.

%-----------------------------------------------------------------------
%-----------------------------------------------------------------------
\section{Conclusions}\label{sec:conc}

In this paper, we proposed DeepSonar that discerns AI-synthesized fake voices by monitoring the learnt neuron behaviors from voice synthesis system.
Overall, our work presents a new insight for detecting AI aided multimedia fakes by monitoring neuron behaviors, which aims to build an effective and robust detector.
Experiments on the three datasets demonstrate its effectiveness and robustness, with potential in the real-world noisy environment.
In fighting against AI-synthesized voices in the wild, robustness should be considered as a priority in designing a detector, since various manipulations on voices can be easily camouflaged as regular operations, while manipulation on images is limited and easy to be spotted. Furthermore, the inconsistency of audio and visual in video DeepFakes is an important clue for DeepFake forensics, thus how to combine recent advances in fake still image and fake voice detection to spot the inconsistency is an important topic for future research. Our neuron behaviors based technique may be a promising idea. Producing and fighting fakes in the AI era is like a mouse and cat game. More powerful techniques should be continuously developed for fighting AI aided fakes as new techniques for producing various fakes will emerge inadvertently.
Our future work would continuously investigate how the proposed DeepSonar method can be extended to or work in tandem with various detectors \cite{huang2020fakelocator,icip16_paint,arxiv20_deeprhythm,wang2019fakespotter,arxiv20_fakepolisher} on other modalities of the `fakes' such as AI-generated / forged images, and DeepFake videos, \etc.

\begin{acks}
This research was supported in part by Singapore National Cybersecurity R\&D Program No. NRF2018NCR-NCR005-0001, National Satellite of Excellence in Trustworthy Software System No. NRF2018NCR-NSOE003-0001, NRF Investigatorship No. NRFI06-2020-0022. It was also supported by JSPS KAKENHI Grant No. 20H04168, 19K24348, 19H04086, and JST-Mirai Program Grant No. JPMJMI18BB, Japan. We gratefully acknowledge the support of NVIDIA AI Tech Center (NVAITC) to our research.
\end{acks}

\balance

% \clearpage
% \newpage
\bibliographystyle{ACM-Reference-Format}
\bibliography{ref}

% \appendix

\end{document}